\RequirePackage{fix-cm}
\documentclass[twocolumn,epjc3]{svjour3}   
\pdfoutput=1 

\usepackage{graphicx}
\usepackage{float}
\usepackage{subfig}
\usepackage{amsmath}
\usepackage{amssymb}
\usepackage{verbatim}
\usepackage{array}
\usepackage{hyperref}
\usepackage{xcolor}
\hypersetup{
    colorlinks,
    linkcolor={red!50!black},
    citecolor={blue!50!black},
    urlcolor={blue!80!black}
}
\usepackage[T1]{fontenc}
\DeclareMathSymbol{\m}{\mathbin}{AMSa}{"39}

\newcommand{\be} {\begin{equation}}
\newcommand{\ee} {\end{equation}}
\newcommand{\om} {\omega}
\newcommand{\ob} {\bar{\omega}}
\newcommand{\ta} {\tau}
\newcommand{\tb} {\bar{\tau}}
\newcommand{\rs} {\boldsymbol{1}}
\newcommand{\rsp} {\boldsymbol{1'}}

\newcommand{\rd} {\boldsymbol{2}}
\newcommand{\rt} {\boldsymbol{3}}
\newcommand{\rtp} {\boldsymbol{3'}}
\newcommand{\rx} {\boldsymbol{6}}
\newcommand{\rxs} {\boldsymbol{\bar{6}}}
\newcommand{\bt} {Y_{24}}
\newcommand{\nr} {N}
\newcommand{\phia} {\acute{\Phi}}
\newcommand{\phib} {\grave{\Phi}}
\newcommand{\ssa} {\acute{S}}
\newcommand{\ssb} {\grave{S}}
\newcommand{\bind} {\Delta}

\newcommand{\off} {\Theta}
\newcommand{\Uac}{U_\text{BM}}
\newcommand{\Ubc}{U_{\theta}}
\newcommand{\Utm} {U_{TM_1}\!({\scriptstyle \frac{-\pi}{12}, \frac{\pi}{2} })}
\newcommand{\Cma} {C_{1}}
\newcommand{\Cmb} {C_{2}}
\newcommand{\Cwa} {C_{\omega1}}
\newcommand{\Cwb} {C_{\omega2}}
\newcommand{\Cwc} {C_{\omega3}}
\newcommand{\Cwd} {C_{\omega4}}
\newcommand{\Cta} {C_{\tau1}}
\newcommand{\Ctb} {C_{\tau2}}
\newcommand{\Ctc} {C_{\tau3}}
\newcommand{\edit} {\color{black}}

\journalname{Eur. Phys. J. Plus}

\begin{document} 

\title{\boldmath $\text{TM}_1$ neutrino mixing with $\sin \theta_{13}=\frac{1}{\sqrt{3}}\sin \frac{\pi}{12}$}
\author{R.~Krishnan\thanksref{e1,addr1}}
\thankstext{e1}{krishnan.rama@saha.ac.in; krisphysics@gmail.com}
\institute{Saha Institute of Nuclear Physics, 1/AF Bidhannagar, Kolkata 700064, India, 
\\https://orcid.org/0000-0002-0707-3267 \label{addr1}}  
\date{Published: 21 April 2022}
\maketitle

\begin{abstract}
	We construct a neutrino model using the flavour group $S_4\times C_4 \times C_3{\edit \times C_2}$ under the type-1 seesaw mechanism. The vacuum alignments of the flavons in the model lead to $\text{TM}_1$ mixing with $\sin \theta_{13}=\frac{1}{\sqrt{3}}\sin \frac{\pi}{12}$. The mixing also exhibits $\mu\text{-}\tau$~reflection symmetry. By fitting the eigenvalues of the effective seesaw mass matrix with the observed neutrino mass-squared differences, we predict the individual light neutrino masses. The vacuum alignment of the $S_4$ triplet appearing in the Majorana mass term plays a key role in obtaining the aforementioned $\text{TM}_1$ scenario. Since the symmetries of the flavour group are not sufficient to define this alignment, we apply the recently proposed framework of the auxiliary group in our model. Using this framework, the $S_4$ triplet is obtained by coupling together several irreducible multiplets that transform under an expanded flavour group consisting of the original flavour group as well as an auxiliary group. The vacuum alignment of each of these multiplets is uniquely defined in terms of its residual symmetries under the expanded flavour group. As a result, the $S_4$ triplet constructed from these multiplets also becomes uniquely defined.
\end{abstract}

\section{Introduction}
\label{sec:intro}

The tribimaximal mixing (TBM)~\cite{Harrison:2002er},
\be\label{eq:tbm}
U_{\text{TBM}} =\left(\begin{matrix} \frac{\sqrt{2}}{\sqrt{3}} &  \frac{1}{\sqrt{3}} & 0 \\[6pt]
	\frac{-1}{\sqrt{6}}  &  \frac{1}{\sqrt{3}} & \frac{1}{\sqrt{2}} \\[6pt]
	\frac{-1}{\sqrt{6}}  & \frac{1}{\sqrt{3}} &  \frac{-1}{\sqrt{2}}
\end{matrix}\right),
\ee
has been the most widely studied mixing ansatz in literature. Though the discovery of nonzero reactor angle has ruled out this ansatz, model builders often use it as a starting point. One approach is to use TBM as the lowest order approximation and then obtain realistic mixing patterns through higher-order corrections. Another approach is to directly obtain realistic ansatze that are modifications of TBM. Such mixing scenarios preserve certain symmetries of TBM while breaking certain others. A nomenclature for the mixing matrix that preserves the $i^{\text{th}}$ column (row) of $U_{\text{TBM}}$ was proposed as $\text{TM}_i$ ($\text{TM}^i$)~\cite{Albright:2008rp,Albright:2010ap}. 

In this paper, we construct a model that results in $\text{TM}_1$ mixing~\cite{Xing:2006ms}. $\text{TM}_1$ preserves the first column of $U_{\text{TBM}}$ and mixes its second and third columns. Therefore, we have
\begin{align}\label{eq:tma}
\begin{split}
U_{\text{TM}_1} (\theta,\zeta) &= U_{\text{TBM}}\cdot\left(\begin{matrix} 1 &  0 & 0 \\
0  & \cos \theta & \sin \theta e^{-i\zeta} \\
0  & -\sin \theta e^{i\zeta} & \cos \theta
\end{matrix}\right)\\[2pt]
&= \left(\begin{matrix} \frac{\sqrt{2}}{\sqrt{3}} &  \frac{\cos \theta}{\sqrt{3}} &  \frac{\sin \theta }{\sqrt{3}}e^{-i\zeta}\\[6pt]
\frac{-1}{\sqrt{6}}  &   \frac{\cos \theta}{\sqrt{3}} - \frac{ \sin \theta}{\sqrt{2}} e^{i\zeta} & \frac{\sin \theta }{\sqrt{3}}e^{-i\zeta}+ \frac{\cos \theta}{\sqrt{2}} \\[6pt]
\frac{-1}{\sqrt{6}}  & \frac{\cos \theta}{\sqrt{3}} + \frac{ \sin \theta}{\sqrt{2}} e^{i\zeta} & \frac{\sin \theta }{\sqrt{3}}e^{-i\zeta}- \frac{\cos \theta}{\sqrt{2}}
\end{matrix}\right).
\end{split}
\end{align}
Several models have been proposed that generate $\text{TM}_1$ mixing. A model with two highly degenerate right-handed neutrinos resulting in resonant leptogenesis and $\text{TM}_1$ mixing is constructed in Ref.~\cite{Xing:2006ms}. The discrete group $S_4$ is often used to implement $\text{TM}_1$ mixing~\cite{Luhn:2013vna,Li:2013jya,Varzielas:2012pa,Shimizu:2017fgu}. In Ref.~\cite{Zhao:2015bza}, a modified Friedberg-Lee symmetry is utilised. A $\text{TM}_1$ scenario with one texture zero is discussed in Ref.~\cite{Gautam:2018izb}. $\text{TM}_1$ scenarios with $\mu\text{-}\tau$~reflection symmetry and maximally broken CP are discussed in Refs.~\cite{Li:2013jya,Rodejohann:2017lre}. Constrained sequential dominance (CSD) models often predict $\text{TM}_1$ mixing along with the lightest neutrino mass being equal to zero. We have several such CSD models implemented using $A_4$~\cite{Antusch:2011ic,King:2013xba,King:2013iva} and $S_4$~\cite{King:2015dvf,King:2016yvg} symmetries.

Using the parametrisation, Eq.~(\ref{eq:tma}), we obtain
\begin{align}
\sin^2 \theta_{13}& = \frac{\sin^2 \theta}{3}\,, \label{eq:theta13}\\
\sin^2 \theta_{12}& = 1-\frac{2}{3-\sin^2 \theta}\,, \label{eq:theta12}\\
\sin^2 \theta_{23}& = \frac{1}{2}\left(1+\frac{\sqrt{6}\sin 2\theta \cos \zeta}{3-\sin^2 \theta}\right)\,, \label{eq:theta23}\\
J& = \frac{\sin 2 \theta \sin \zeta}{6 \sqrt{6}}, \label{eq:jcp}
\end{align}
where 
\begin{align}\label{eq:deltacp}
\begin{split}
J&=\text{Im}(U_{\mu3}U^*_{e3}U_{e2}U^*_{\mu2})\\
&=\frac{1}{8}\sin \delta \sin 2\theta_{12}\sin 2\theta_{23}\sin 2\theta_{13}\cos \theta_{13}
\end{split}
\end{align}
is the Jarlkog's rephasing invariant~\cite{Jarlskog:1985ht,Jarlskog:1985cw,Jarlskog:1986ia,Jarlskog:1987wq,Jarlskog:2004be}.

Global fit~\cite{Esteban:2018azc} of neutrino oscillation data gives 
\begin{align}
\sin^2\theta_{13}&=0.02241^{+0.00066}_{-0.00065}\,.\label{eq:gf13}\\
\sin^2\theta_{12}&=0.310^{+0.013}_{-0.012}\,,\label{eq:gf12}
\end{align}
This leads to $|U_{12}|^2 = \sin^2\theta_{12} \cos^2\theta_{13} =0.303^{+0.012}_{-0.012}$ and $|U_{12}|^2+|U_{13}|^2 = \sin^2\theta_{12} \cos^2\theta_{13}+\sin^2\theta_{13}\,\,=\,\,$ $0.325^{+0.013}_{-0.012}$. The TBM prediction of $|U_{12}|^2 = \frac{1}{3}$ is disfavoured while the $\text{TM}_1$ constraint $|U_{12}|^2+|U_{13}|^2 = \frac{1}{3}$ is consistent with the experimental fit at $1\sigma$ level. Therefore, we may argue that the most promising among $\text{TM}_i$ ($\text{TM}^i$) mixings is perhaps $\text{TM}_1$. 

Discrete groups have been used extensively to model the observed flavour symmetries~\cite{1003.3552,Altarelli:2010gt,Grimus:2011mp,1204.0445,Ishimori:2012zz,King:2013eh,1402.4271,1701.04413,Meloni:2017cig,1711.10806}. The three families of fermions are often assumed to transform as a triplet under a given discrete group. Flavour models may also include scalar fields called flavons which transform as various multiplets under the group. The Yukawa couplings and thus the fermion mass matrices in the Standard Model (SM) are obtained in terms of the Vacuum Expectation Values (VEVs) of these flavon fields. The VEVs emerge through Spontaneous Symmetry Breaking (SSB) of the flavon potentials. Therefore, the features of the mass matrices have their origin in the properties of the flavour group such as the types of its irreducible multiplets, the Clebsch-Gordan (C-G) coefficients appearing in the tensor products of these multiplets, the alignment of the flavon VEVs etc. These general principles underlie the model presented in our paper also.

The paper can be broadly divided into two parts. The first part consists of Sections~\ref{sec:s4}-\ref{sec:phen} where a type-1 seesaw model is constructed using the discrete group $S_4$.  In Section~\ref{sec:s4}, we briefly mention the essential features of the $S_4$ group and its representations. In Section~\ref{sec:model}, we introduce the flavon content of the model which includes a triplet ($\chi$) in the charged-lepton sector and a singlet ($s$) and a triplet ($\phi$) in the Majorana neutrino sector. We construct the mass terms using these flavons, the SM fields and the right-handed neutrinos. We also construct the potential terms of the flavons and obtain their VEVs through the mechanism of SSB. The charged-lepton and the neutrino mass matrices are obtained in terms of these VEVs. Section~\ref{sec:phen} covers the phenomenology where we extract the PMNS mixing parameters as well as the light neutrino masses. We compare the model's predictions with the experimental data. The symmetries and the features of $S_4$ as well as the alignments of the flavon VEVs lead to the results predicted by the model. Notably, the result $\sin\theta_{13}=\frac{1}{\sqrt{3}}\sin \frac{\pi}{12}$ is obtained as the consequence of a specific VEV chosen for the triplet flavon, $\phi$. 

In the second part of the paper, consisting of Sections~\ref{sec:defvev}-\ref{sec:recast}, we discuss the problem of obtaining the alignments of the VEVs based on discrete symmetries. In Section~\ref{sec:defvev}, we study all possible alignments of the triplets of $S_4$ that can be uniquely defined in terms of their residual symmetries alone. We find that the residual symmetries under $S_4$ are not sufficient to define the vacuum alignment of the triplet $\phi$ that we mentioned earlier. In this context, we briefly describe the recently proposed framework of the auxiliary group~\cite{Krishnan:2019ftw,2011.11653} with which additional symmetries can be incorporated to obtain a richer choice of vacuum alignments. To apply this framework to our model, we construct a discrete group (which we call $Y_{24}$) with suitable symmetries in Section~\ref{sec:aux}. We use this group in Section~\ref{sec:recast} to recast the model in the framework of the auxiliary group. Here, the triplet $\phi$ is replaced with an `effective' triplet consisting of $\phia$, $\phib$ and $\Delta$ which are `elementary' flavons transforming as multiplets of the expanded flavour group in our framework. Their VEVs are fully defined in terms of their respective residual symmetries and thus the VEV of the effective triplet also becomes uniquely defined. We finally conclude in Section~\ref{sec:conclusion}.


\section{The Group $S_4$}
\label{sec:s4}

The group $S_4$ has been used extensively in neutrino models~\cite{Luhn:2013vna,Varzielas:2012pa,Shimizu:2017fgu,King:2015dvf,King:2016yvg,Brown:1984dk,Lee:1994qx,Mohapatra:2003tw,Ma:2005pd,Hagedorn:2006ug,Zhang:2006fv,Caravaglios:2006aq,Koide:2007sr,Krishnan:2012me,Krishnan:2012sb,2003.00506}. $S_4$ is the group of permutations of four objects. We may define it using the presentation,
\be
\langle P, Q, R ~|~ P^2 =Q^3 = R^4 = PQR = I \rangle.
\ee
Therefore, $S_4$ is the von Dyck group with parameters (2,3,4). As can be inferred from its presentation, only two of its generators are independent, i.e.
\be
P=QR,\quad Q=PR^3,\quad R=Q^2P.
\ee
The conjugacy classes and the irreducible representations of $S_4$ are listed in Table~\ref{tab:char}. 

{\renewcommand{\arraystretch}{1.2}
	\begin{table}[tbp]
		\begin{center}
			\begin{tabular}{|c|c c c c c|}
				\hline
				&$()$ & $(12)(34)$	& $(12)$ & $(1234)$ & $(123)$\\
				\hline
				$\rs$ & $1$ & $1$ & $1$ & $1$ & $1$ \\
				$\rsp$ & $1$ & $1$ & $-1$ & $-1$ & $1$  \\
				$\rd$ & $2$ & $2$ & $0$ & $0$ & $-1$ \\
				$\rt$ & $3$ & $-1$ & $-1$ & $1$& $0$ \\
				$\rtp$ & $3$ & $-1$ & $1$ & $-1$& $0$ \\
				\hline
			\end{tabular}
		\end{center}
		\caption{The character table of $S_4$.}
		\label{tab:char}
\end{table}}

\begin{figure}
	\begin{tabular}{cc}
		\includegraphics[width=50mm,trim={2.0cm 4.5cm 0.5cm 1.0cm},clip]{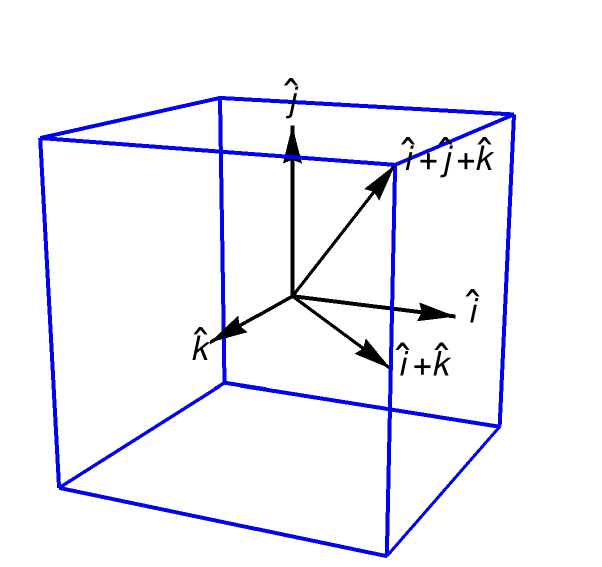}&\includegraphics[width=32mm,trim={1.2cm 0.8cm 0.5cm 0.9cm},clip]{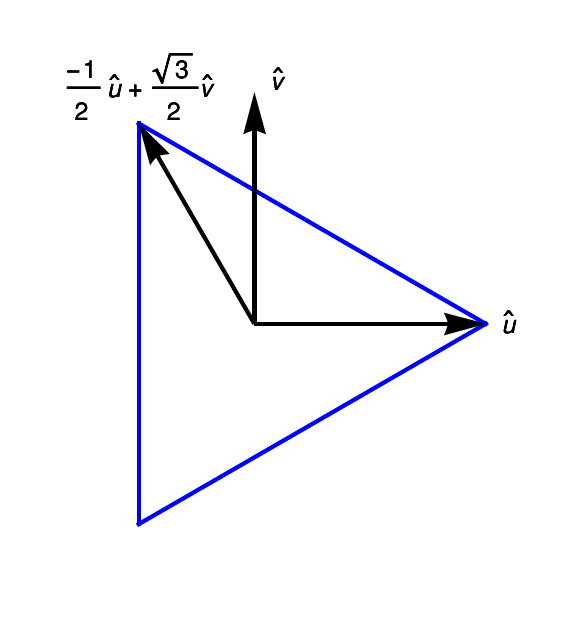} \\
		(a) & (b) \\[6pt]
	\end{tabular}
	\caption{The basis states and the symmetries of the representations $\rt$ and $\rd$ of $S_4$.}
	\label{fig:basis}
\end{figure}

The defining triplet representation, $\rt$, can be generated using the matrices,
\be\label{eq:gens}
P =  \left(\begin{matrix} 0 & 0 & 1\\
	0 & -1 & 0 \\
	1 & 0 & 0
\end{matrix}\right),\,\, Q =  \left(\begin{matrix} 0 & 1 & 0\\
	0 & 0 & 1 \\
	1 & 0 & 0
\end{matrix}\right),\,\, R =  \left(\begin{matrix} 1 & 0 & 0\\
	0 & 0 & 1 \\
	0 & -1 & 0
\end{matrix}\right).
\ee
This triplet representation denotes the rotational symmetries of a cube, Figure~\ref{fig:basis}(a). In the figure, the unit vectors $\hat{i}$, $\hat{j}$ and $\hat{k}$ correspond to the basis states $(1,0,0)^T$, $(0,1,0)^T$ and $(0,0,1)^T$ respectively. $P$, $Q$ and $R$ are rotations (clockwise) by angles $\pi$, $\frac{2\pi}{3}$ and $\frac{\pi}{2}$ about the axes aligned along the directions $\hat{i}+\hat{k}$, $\hat{i}+\hat{j}+\hat{k}$ and $\hat{i}$ respectively. $S_4$ has 24 elements which fall under five conjugacy classes, Table~\ref{tab:char}. The class $(123)$ consists of the elements conjugate to $Q$. They represent rotations by angles $\frac{2\pi}{3}$ and $\frac{4\pi}{3}$ about the body diagonals of the cube. There are 8 elements in this conjugacy class. The elements in the class $(12)$ are those that are conjugate to $P$. They represent $\pi$-rotations about the axes passing through the centres of the opposite edges of the cube. We have 6 such rotations. The conjugacy class $(1234)$ also has 6 elements. They are conjugate to $R$ and they represent rotations by angles $\frac{\pi}{2}$ and $\frac{3\pi}{2}$ about the axes aligned along the basis states. On the other hand, $\pi$-rotations about these axes constitute the conjugacy class $(12)(34)$. We have 3 such rotations and they are conjugate to $R^2$.

In model building, a commonly adopted complex basis for $\rt$ involves the generators $S$, $T$ and $U$,
\begin{align}\label{eq:complexbases}
\begin{split}
&S = \frac{1}{3}\left(\begin{matrix}  -1 &  2 & 2\\
2 & -1 & 2 \\
2 & 2 &  -1
\end{matrix}\right), \quad T =  \left(\begin{matrix} 1 & 0 & 0\\
0 & \ob & 0 \\
0 & 0 & \om
\end{matrix}\right),\\[2pt] & \quad \quad \quad \quad\quad \quad U =  -\left(\begin{matrix} 1 & 0 & 0\\
0 & 0 & 1 \\
0 & 1 & 0
\end{matrix}\right),
\end{split}
\end{align}
where $\om=e^{i\frac{2\pi}{3}}$ and $\ob=e^{-i\frac{2\pi}{3}}$ are the complex cube roots of unity. The two sets of bases, Eqs.(\ref{eq:gens}, \ref{eq:complexbases}), are related by
\begin{align}
S&=U_\om \, R^2\,U_\om^\dagger,\\
T&=U_\om \,Q\,U_\om^\dagger,\\
U&=U_\om \,PRQP\,U_\om^\dagger=PRQP,
\end{align}
where
\be\label{eq:tm}
U_\om=
\frac{1}{\sqrt{3}}\left(\begin{matrix}1 & 1 & 1\\
	1 & \om & \ob\\
	1 & \ob & \om
\end{matrix}\right)
\ee
is the $3\times3$ trimaximal matrix. Unlike $P$, $Q$ and $R$, the generators $S$, $T$ and $U$ are independent. $S_4$ can not be generated from any two of them. In this paper, we adopt the basis, Eqs.(\ref{eq:gens}), so as to provide a geometric interpretation to our analysis. 

Besides the defining triplet $\rt$, $S_4$ has another triplet $\rtp$. $\rtp$ differs from $\rt$  by having a space inversion (multiplication by $-1$) in addition to the usual rotations for the elements of the conjugacy classes $(12)$ and $(1234)$ . We have
\be\label{eq:tripletpgens}
P(\rtp) = -P, \quad Q(\rtp) = Q, \quad R(\rtp) = -R.
\ee
Like $\rt$, $\rtp$ is also a faithful representation. The singlet $\rsp$ involves multiplication with $-1$ corresponding to the space inversion in $\rtp$, i.e.
\be
P(\rsp) = -1, \quad Q(\rsp) = 1, \quad R(\rsp) = -1.
\ee

The generators of the doublet, $\rd$, are
\begin{align}\label{eq:doubletgens}
\begin{split}
P(\rd) &=  \left(\begin{matrix} -\frac{1}{2} & -\frac{\sqrt{3}}{2}\\[4pt]
-\frac{\sqrt{3}}{2} & \frac{1}{2}
\end{matrix}\right), \quad Q(\rd) = \left(\begin{matrix} -\frac{1}{2} & \frac{\sqrt{3}}{2}\\[4pt]
-\frac{\sqrt{3}}{2} & -\frac{1}{2} 
\end{matrix}\right),\\[2pt] &\quad\quad\quad\quad \quad R(\rd) =  \left(\begin{matrix} 1 & 0\\
0 & -1
\end{matrix}\right).
\end{split}
\end{align}
$\rd$ forms the symmetry group of an equilateral triangle, Figure~\ref{fig:basis}(b). $P(\rd)$ and $R(\rd)$ are the reflections about the directions $\frac{-1}{2}\hat{u}+\frac{\sqrt{3}}{2}\hat{v}$ and $\hat{u}$ respectively where $\hat{u}$ and $\hat{v}$ are the basis states $(1,0)^T$ and $(0,1)^T$ respectively. $Q(\rd)$ is the rotation (clockwise) by an angle $\frac{2\pi}{3}$ in the $\hat{u}$-$\hat{v}$ plane.

The tensor product expansion of two triplets ($\rtp$s) is given by
\be
\rtp\times\rtp=\rs+\rd+\rtp+\rt.
\ee
With $a=(a_1, a_2, a_3)$ and $b=(b_1, b_2, b_3)$ transforming as $\rtp$s, we obtain
\begin{align}
\left(ab\right)_{\rs}& = a_1 b_1 + a_2 b_2 + a_3 b_3\,, \label{eq:tp1}\\
\left(ab\right)_{\rd}& = \left(2 a_1 b_1 - a_2 b_2 - a_3 b_3, \sqrt{3} (a_2 b_2 - a_3 b_3)\right)^T\,, \label{eq:tp2}\\
\left(ab\right)_{\rtp}& = (a_2 b_3+a_3 b_2, a_3 b_1+a_1 b_3, a_1 b_2+a_2 b_1)^T\,, \label{eq:tp3p}\\
\left(ab\right)_{\rt}& = (a_2 b_3-a_3 b_2, a_3 b_1-a_1 b_3, a_1 b_2-a_2 b_1)^T\,, \label{eq:tp3}
\end{align}
where $\left(ab\right)_{\rs}$, $\left(ab\right)_{\rd}$, $\left(ab\right)_{\rtp}$ and $\left(ab\right)_{\rt}$ denote the irreducible parts in the tensor product of $a$ and $b$ that transform as $\rs$, $\rd$, $\rtp$ and $\rt$ respectively. The tensor product of two doublets, $p=(p_1, p_2)$ and $q=(q_1, q_2)$, leads to
\be
\rd\times\rd=\rs+\rsp+\rd,
\ee
with
\begin{align}
\left(pq\right)_{\rs}& = p_1 q_1 + p_2 q_2\,, \label{eq:db1}\\
\left(pq\right)_{\rsp}& = p_1 q_2 - p_2 q_1\,, \label{eq:db1p}\\
\left(pq\right)_{\rd}& = (-p_1 q_1 + p_2 q_2, p_1 q_2 + p_2 q_1)^T\,. \label{eq:db2}
\end{align}


\section{The model constructed using $S_4\times C_4\times C_3 \,{\edit \times \, C_2}$}
\label{sec:model}

{\renewcommand{\arraystretch}{1.4}
	\setlength{\tabcolsep}{5pt}
	\begin{table}[tbp]
		\begin{center}
			\begin{tabular}{|c|c c c c c c c c c c|}
				\hline
				&$L$ & $e_R$	&$\mu_R$&$\tau_R$	&$\nr$&$\edit \rho$&$\chi$&$s$&$\phi$&$\eta$\\
				\hline
				$S_4$ &$\edit \rtp$ & $\rs$	&$\rs$&$\rs$&$\edit \rtp$&$\edit \rs$&$\edit \rtp$&$\rs$&$\rtp$&$\rd$\\
$C_4\times C_3$ &$i$ & $i$ &$i \om$&$i \ob$&$i$&$\edit1$&$\om$&$-1$&$-1$&$1$\\
$\edit C_2$ &$\edit -1$ & $\edit 1$ &$\edit 1$&$\edit -1$&$\edit 1$&$\edit -1$&$\edit 1$&$\edit 1$&$\edit 1$&$\edit 1$\\
				\hline
			\end{tabular}
		\end{center}
		\caption{The fields in the model as the multiplets under $S_4\times C_4\times C_3{\edit \, \times\, C_2}$. The standard model Higgs ($H$) is assumed to remain invariant under the flavour group.}
		\label{tab:flavourcontent}
\end{table}}

The field content of the model is given in Table~\ref{tab:flavourcontent}. The three families of the left-handed weak-isospin lepton doublets and the three right-handed heavy neutrinos form the $S_4$ triplets, $L$ and $\nr$, respectively. $H$ is the SM Higgs. The flavons, $s$, $\chi$, $\phi$, $\eta$ and $\edit \rho$ are scalar fields and are gauge invariants. Using this field content and their symmetry properties, we obtain the following Lagrangian:
\begin{align}\label{eq:lagr}
\begin{split}
{\mathcal L}=&y_\tau \bar{L} \frac{\chi}{\Lambda} \tau_R H+{\edit y_\mu \bar{L} \frac{\rho \chi^*}{\Lambda^2} \mu_R H + y_e \bar{L} \frac{\rho(\chi^*\chi)_{\rtp}}{\Lambda^3} e_R H }\\
& {\edit+ y_\nu \bar{L} \nr \frac{\rho}{\Lambda}\widetilde{H} + y_s \left( \bar{\nr^c} \nr \right)_{\rs}  s + y_\phi \left( \bar{\nr^c} \nr \right)_{\rtp} \phi},
\end{split}
\end{align}
where $\Lambda$ is the cut-off scale of the theory and $y_x$ are the Yukawa-like coupling constants. {\edit Along with the symmetries given Table~\ref{tab:flavourcontent}, we also impose CP symmetry. This implies that the coupling constants $y_x$ are real.} In Eq.~(\ref{eq:lagr}), $\left(\right)_{\rs}$, $\left(\right)_{\rt}$ and $\left(\right)_{\rtp}$ denote the tensor products that transform as $\rs$, $\rt$ and $\rtp$ respectively under $S_4$.

The flavon $\chi$ transforms as $\rtp$ under $S_4$ and $\om$ under $C_3$. We have introduced the $C_3$ group so that $\chi$, $\chi^*$ and $(\chi^*\chi)_{\rtp}$ couple with $\tau_R$, $\mu_R$ and $e$ respectively along with the $S_4$ triplet $L$. They form the mass terms in the charged-lepton sector. {\edit Note that $\chi$ is the only  flavon that is complex, and its complex conjugation corresponds to charge conjugation. The VEV of $\chi$ spontaneously breaks CP symmetry, and it is the only source of CP violation in the model. The flavon $\rho$ and its $C_2$ charge aids in creating the observed charged-lepton mass hierarchy through the Froggatt-Nielsen mechanism.}

The next term in the Lagrangian, $\edit y_\nu \bar{L} \nr \frac{\rho}{\Lambda}\widetilde{H}$, is the Dirac mass term for the neutrinos. The flavons, $s$ and $\phi$, transform as $\rs$ and $\rtp$ and couple with $\left( \bar{\nr^c} \nr \right)_{\rs}$ and $\left( \bar{\nr^c} \nr \right)_{\rtp}$ respectively. They constitute the Majorana mass term for the neutrinos. We have introduced the $C_4$ group to prevent unwanted couplings. {\edit The flavour symmetry breaking scale is expected to be very high, typically the grand unification scale.} Through the type-1 seesaw mechanism, the light-neutrino masses get suppressed by this scale. 

The Higgs acquires the VEV,
\be \label{eq:vevh}
\langle H\rangle = (0,v)^T, 
\ee
through SSB. The VEV corresponds to a minimum point in the Higgs potential. The flavon fields also acquire VEVs through SSB,
\begin{align}
{\edit \langle \rho \rangle}&{\edit= v_\rho,}\label{eq:vevrho}\\
\langle\chi\rangle &=v_{\chi} (1,\om,\ob)^T,\label{eq:vevchi}\\
\langle s\rangle&= v_s,\label{eq:vevs}\\
\langle\eta\rangle &= v_{\eta} \left(-\frac{1}{2},\frac{\sqrt{3}}{2}\right)^T,\label{eq:veveta}\\
\langle\phi\rangle &=v_{\phi} \left(-\frac{1}{2\sqrt{2}},-\sqrt{3},\frac{1}{2\sqrt{2}}\right)^T.\label{eq:vevphi} 
\end{align}
In the following section, we construct the flavon potentials that lead to these VEVs. Note that the flavon $\eta$ does not couple to any of the fermions in the Lagrangian at the lowest order and hence it does not contribute directly to the fermion mass matrices. We use it as a driving field~\cite{Altarelli:2005yx} in the construction of the potential of the flavon $\phi$.


\subsection{The flavon potentials}
\label{subsec:potentials}

\subsubsection*{The charged-lepton sector ($\chi$)}

{\edit
For the singlet flavon $\rho$, we write the potential,
\be
\mathcal{V}_\rho =  k_\rho \rho^4 - 2 k_\rho v_\rho^2 \rho^2,
\ee
where $k_\rho$ is a real dimensionless parameter, and $v_\rho$ is a real parameter of mass dimension one. For $k_\rho>0$, this potential has two points of minima,
\be
\rho=\pm v_\rho.
\ee
Through SSB, we obtain one of these minima as the VEV, 
\be
\langle \rho\rangle= v_\rho.
\ee}
With $\chi=(\chi_1, \chi_2, \chi_3)$  and using Eqs.~(\ref{eq:tp1}-\ref{eq:tp3p}), we construct the following quadratic expressions:
\be
\left(\chi^*\chi\right)_{\rs} = \chi_1^* \chi_1 + \chi_2^* \chi_2 + \chi_3^* \chi_3\,, \label{eq:pottp1s}
\ee
\begin{align}
\left(\chi\chi\right)_{\rs}& = \chi_1^2 + \chi_2^2 + \chi_3^2\,,\\
\left(\chi\chi\right)_{\rd}& = \left(2 \chi_1^2 - \chi_2^2 - \chi_3^2, \sqrt{3} (\chi_2^2 - \chi_3^2)\right)^T\,,\\
\left(\chi\chi\right)_{\rtp}& = 2 \left(\chi_2 \chi_3, \chi_3 \chi_1, \chi_1 \chi_2\right)^T\,,
\end{align}
The antisymmetric expression $\left(\chi\chi\right)_{\rt}$ vanishes. 
At the quadratic order, the only invariant term is 
\be
{\mathcal{T}(\chi^2)} = \left(\chi^*\chi\right)_{\rs}.
\ee
{\edit At the cubic order, we have the invariant,
\be
{\mathcal{T}(\chi^3)} = \text{Re}[\left(\chi\chi\right)_{\rtp}^T\chi]=6\,\text{Re}[\chi_1\chi_2\chi_3]
\ee}
 At the quartic order, we obtain the following invariants:
\begin{align}
{\mathcal{T}_1(\chi^4)} &= \left(\chi\chi\right)_{\rs}^* \left(\chi\chi\right)_{\rs}, \\
{\mathcal{T}_2(\chi^4)} &= \left(\chi\chi\right)_{\rd}^\dagger\left(\chi\chi\right)_{\rd},\\
{\mathcal{T}_3(\chi^4)} &= \left(\chi\chi\right)_{\rtp}^\dagger\left(\chi\chi\right)_{\rtp}.
\end{align}
${\mathcal{T}(\chi^2)}^2$ is related to the above invariants through the relation,
\be
6({\mathcal{T}(\chi^2)})^2 = 2{\mathcal{T}_1(\chi^4)}+{\mathcal{T}_2(\chi^4)}+3{\mathcal{T}_3(\chi^4)}.
\ee
Hence we have only three independent invariants at the quartic order. We use them to construct the flavon potential,
\begin{align}
\begin{split}
\mathcal{V}_\chi = &{\edit \kappa_{0} v_\chi {\mathcal{T}(\chi^3)}+}\kappa_{1} {\mathcal{T}_1(\chi^4)}+ \kappa_{2} {\mathcal{T}_2(\chi^4)}+ \kappa_{3} {\mathcal{T}_3(\chi^4)}\\
&-({\edit 3\kappa_0+}12\kappa_2+8\kappa_3) v_\chi^2 \, {\mathcal{T}(\chi^2)},
\end{split}
\end{align}
where ${\edit \kappa_0}, \kappa_1, \kappa_2, \kappa_3$ are dimensionless real parameters and $v_\chi$ is a real parameter of mass dimension one. By calculating the first derivative of this potential with respect to the components of $\chi$, we can show that it has a set of extremum points,
\be
\chi=g_i  \, v_{\chi} (1,\om,\ob)^T,
\ee
where $g_i$ are the elements of the group $S_4\times C_3$. Through SSB, the flavon acquires one among these extrema as its VEV~\footnote{For a large region of the parameter space $({\edit \kappa_0}, \kappa_1, \kappa_2, \kappa_3, v_\chi)$, the extrema correspond to the minima. For the stability of the VEV, we assume that the parameters fall in this region.}, 
\be
\langle\chi\rangle=v_{\chi} (1,\om,\ob)^T.
\ee


\subsubsection*{The neutrino sector ($s$, $\eta$, $\phi$)}
Construction of potential for the singlet $s$ is similar to that of $\rho$, i.e.
\be
\mathcal{V}_s =  k_s s^4 - 2 k_s v_s^2 s^2,
\ee
resulting in the VEV
\be
\langle s\rangle= v_s.
\ee

We use Eqs.~(\ref{eq:db1}, \ref{eq:db2}) to obtain the tensor product of two flavon doublets, $\eta=(\eta_1,\eta_2)$, resulting in the quadratic invariant,
\be
{\mathcal{T}(\eta^2)} =\left(\eta\eta\right)_{\rs}= \eta_1^2 + \eta_2^2\,,
\ee
and the doublet,
\be
\left(\eta\eta\right)_{\rd}= \left(-\eta_1^2 + \eta_2^2, 2\eta_1\eta_2\right)^T\,.
\ee
The antisymmetric expression $\left(\eta\eta\right)_{\rsp}$ vanishes. Using $\left(\eta\eta\right)_{\rd}$, we obtain the cubic invariant,
\be
{\mathcal{T}(\eta^3)} = \eta^T  \left(\eta\eta\right)_{\rd}=-\eta_1^3+3\eta_1\eta_2^2.
\ee
At the quartic order, we have the invariant
\be
{\mathcal{T}(\eta^4)} = \left(\eta\eta\right)_{\rd}^T \left(\eta\eta\right)_{\rd} = (\eta_1^2 + \eta_2^2)^2 .
\ee
Note that ${\mathcal{T}(\eta^4)}=({\mathcal{T}(\eta^2)})^2$. 

Using the flavon $\phi$, we obtain the following quadratic expressions:
\begin{align}
\left(\phi\phi\right)_{\rs}& = \phi_1^2 + \phi_2^2 + \phi_3^2\,, \label{eq:phis}\\
\left(\phi\phi\right)_{\rd}& = \left(2 \phi_1^2 - \phi_2^2 - \phi_3^2, \sqrt{3} (\phi_2^2 - \phi_3^2)\right)^T\,, \label{eq:phid}\\
\left(\phi\phi\right)_{\rtp}& = 2 \left(\phi_2 \phi_3, \phi_3 \phi_1, \phi_1 \phi_2\right)^T\,, \label{eq:phit}
\end{align}
with 
\be
{\mathcal{T}(\phi^2)} = \left(\phi\phi\right)_{\rs} 
\ee
being the quadratic invariant. At the quartic order, we obtain the invariants,
\begin{align}
{\mathcal{T}_1(\phi^4)} &= \left(\phi\phi\right)_{\rs}^2, \\
{\mathcal{T}_2(\phi^4)} &= \left(\phi\phi\right)_{\rd}^T\left(\phi\phi\right)_{\rd},\\
{\mathcal{T}_3(\phi^4)} &= \left(\phi\phi\right)_{\rtp}^T\left(\phi\phi\right)_{\rtp}.
\end{align}
These invariants are related by 
\be
4{\mathcal{T}_1(\phi^4)}= {\mathcal{T}_2(\phi^4)}+3{\mathcal{T}_3(\phi^4)},
\ee
so only two of the quartic invariants are independent.

We can also couple $\eta$ and $\phi$ to obtain the following invariants:
\begin{align}
{\mathcal{T}(\eta\phi^2)} &= \eta^T\left(\phi\phi\right)_{\rd}, \\
{\mathcal{T}(\eta^2\phi^2)} &= \left(\eta\eta\right)_{\rd}^T\left(\phi\phi\right)_{\rd}.
\end{align}
As a result, we obtain a total of eight independent invariant terms involving $\eta$ and $\phi$: ${\mathcal{T}(\eta^2)}$, ${\mathcal{T}(\eta^3)}$, ${\mathcal{T}(\eta^4)}$, ${\mathcal{T}(\phi^2)}$, ${\mathcal{T}_1(\phi^4)}$, ${\mathcal{T}_2(\phi^4)}$, ${\mathcal{T}(\eta\phi^2)}$, ${\mathcal{T}(\eta^2\phi^2)}$. 

Using these terms, we construct the potential,
\begin{align}\label{eq:phipot0}
\begin{split}
\mathcal{V}_{\eta\phi} = &c_1 {\mathcal{T}(\eta^2)}+c_2{\mathcal{T}(\eta^3)}+c_3{\mathcal{T}(\eta^4)}+c_4{\mathcal{T}(\phi^2)}\\&+c_5{\mathcal{T}_1(\phi^4)}+c_6{\mathcal{T}_2(\phi^4)}+c_7{\mathcal{T}(\eta\phi^2)}+c_8{\mathcal{T}(\eta^2\phi^2)}.
\end{split}
\end{align}
These terms can be rearranged to obtain
\begin{align}\label{eq:phipot}
\begin{split}
\mathcal{V}_{\eta\phi} =&k_1 \left({\mathcal{T}(\eta^2)}-v_\eta^2\right)^2+k_2 \left(3{\mathcal{T}(\phi^2)}-(k_c v_\eta^2+4 v_\phi^2)\right)^2\\
&+ k_3 \left(v_\eta \eta+\left(\eta\eta\right)_{\rd}\right)^T\left(v_\eta \eta+\left(\eta\eta\right)_{\rd}\right)  \\
& + k_4 \left(k_c v_\eta \eta-\left(\phi\phi\right)_{\rd}\right)^T\left(k_c v_\eta \eta-\left(\phi\phi\right)_{\rd}\right)\\
&+k_5\left(k_c \left(\eta\eta\right)_{\rd}+\left(\phi\phi\right)_{\rd}\right)^T\left(k_c \left(\eta\eta\right)_{\rd}+\left(\phi\phi\right)_{\rd}\right)\\
&-k_1 v_\eta^4 - k_2 (k_c v_\eta^2+4 v_\phi^2)^2,
\end{split}
\end{align}
where $v_\eta$ and $v_\phi$ are parameters of mass dimension one. The parameters  $k_1, ..., k_5$ and $k_c$ are dimensionless; $k_1, ..., k_5$ are assumed to be positive. The eight parameters, $v_\eta$, $v_\phi$, $k_1, ..., k_5, k_c$ are related to $c_1, ..., c_8$ through the equations,
\begin{align}
c_1 &=   (-2k_1+k_3 +k_4 k_c^2) \, v_\eta^2, &c_5 & = 9 k_2,\\
c_2 &= 2 k_3 v_\eta, &c_6 & = k_4+k_5,\\
c_3 &= k_1 + k_3 +  k_5 k_c^2, &c_7 & = -2 k_4 k_c v_\eta,\\
c_4 &= -6 k_2 (k_c v_\eta^2+4v_\phi^2), &c_8 & = 2 k_5 k_c.
\end{align}
The potential, Eq.~(\ref{eq:phipot}), consists of five terms with the coefficients $k_1$ to $k_5$. Each of them is positive semidefinite and hence they should all vanish when minimised. The constant factor, $-k_1 v_\eta^4 - k_2 (k_c v_\eta^2+4 v_\phi^2)^2$, does not play any role in the minimisation of the potential; it is added only for equating Eq.~(\ref{eq:phipot0}) with Eq.~(\ref{eq:phipot}).  
The first term, $\left({\mathcal{T}(\eta^2)}-v_\eta^2\right)^2$, vanishes when $|\eta|^2 = v_\eta^2$. Its minimum corresponds to a continuous set of points. The third term, $\left(\left(\eta\eta\right)_{\rd}+v_\eta \eta\right)^T\left(\left(\eta\eta\right)_{\rd}+v_\eta \eta\right)$, breaks this continuous symmetry. This term vanishes for 
\be\label{eq:etamin}
\eta= g_i  \, v_\eta \left(-\frac{1}{2},\frac{\sqrt{3}}{2}\right)^T,
\ee
where $g_i$ are the elements of the doublet representation of the group $S_4$. 

The second term, $\left(3{\mathcal{T}(\phi^2)}-(k_c v_\eta^2+4 v_\phi^2)\right)^2$, is invariant under the three-dimensional orthogonal transformation, $O(3)$, of the flavon $\phi$, and this term vanishes when $3|\phi|^2 = k_c v_\eta^2+4 v_\phi^2$. This minimum represents a continuous set of points corresponding to the $O(3)$ symmetry. The fourth and the fifth terms couple $\phi$ with $\eta$ and they break this continuous symmetry to $S_4$. These terms vanish when 
\be\label{eq:phimin}
\phi= g_i  \, v_\phi \left(\frac{1}{\sqrt{2}}\sin \alpha,-2\cos \alpha,-\frac{1}{\sqrt{2}}\sin \alpha\right)^T,
\ee
where 
\be\label{eq:alphaf}
\sin \alpha = \frac{1}{3} \sqrt{8-k_c \frac{v_\eta^2}{v_\phi^2}}, \quad \cos \alpha = \frac{1}{3} \sqrt{1+k_c \frac{v_\eta^2}{v_\phi^2}}
\ee
and $g_i$ are the elements of the triplet representation ($\rtp$) of $S_4$. As a result, we obtain a discrete set of minima for the potential.

Through SSB, the flavons acquire one among the minima, Eqs.~(\ref{eq:etamin}-\ref{eq:phimin}), as their VEVs, 
\begin{align}
\langle\eta\rangle&= v_\eta \left(-\frac{1}{2},\frac{\sqrt{3}}{2}\right)^T,\\
\langle\phi\rangle&=v_\phi \left(\frac{1}{\sqrt{2}}\sin \alpha,-2\cos \alpha,-\frac{1}{\sqrt{2}}\sin \alpha\right)^T.\label{eq:vevalpha}
\end{align}
The VEV assumed in our model, Eq.~(\ref{eq:vevphi}), corresponds to Eq.~(\ref{eq:vevalpha}) with $\alpha=-\frac{\pi}{6}$. We may assign specific values to the parameters in the potential, $v_\phi$, $v_\eta$, $k_c$, so that, using Eq.~(\ref{eq:alphaf}), we obtain $\alpha=-\frac{\pi}{6}$ and thus obtain the desired VEV. However, tuning the parameters in the potential to obtain a specific VEV is not justified if we expect that the underlying discrete symmetries naturally determine the alignment of the VEV. On the other hand, $\alpha$ being equal to $-\frac{\pi}{6}$ may point towards additional symmetries. This question is addressed in the second part of this paper where we use the framework of the auxiliary group to uniquely define the VEV in terms of these additional symmetries.

In the following section, we obtain the mass matrices in the charged-lepton and the neutrino sectors in terms of the VEVs, Eqs.~(\ref{eq:vevh}-\ref{eq:vevchi}, \ref{eq:vevphi}).


\subsection{The charged-lepton mass matrix}
\label{subsec:cl}
The flavon $\chi$ couples in the charged-lepton mass term,
\be \label{eq:masstermc}
y_\tau \bar{L} \frac{\chi}{\Lambda} \tau_R H+{\edit y_\mu \bar{L} \frac{\rho \chi^*}{\Lambda^2} \mu_R H + y_e \bar{L} \frac{\rho(\chi^*\chi)_{\rtp}}{\Lambda^3} e_R H}.
\ee
Due to the $C_3$ assignments in Table~\ref{tab:flavourcontent}, we can see that $\chi$ couples with $\tau_R$, $\chi^*$ couples with $\mu_R$ and $(\chi^*\chi)_{\rt}$ couples with $e_R$ at the lowest order. Using Eq.~(\ref{eq:tp3}), we obtain 
\be\label{eq:chichis}
\edit (\chi^*\chi)_{\rtp} = (\chi_2^* \chi_3 + \chi_3^* \chi_2, \chi_3^* \chi_1 + \chi_1^* \chi_3, \chi_1^* \chi_2 + \chi_2^* \chi_1).
\ee
The VEV of $\chi$, Eq.~(\ref{eq:vevchi}), and its conjugate,
\be\label{eq:chivevs}
\langle\chi\rangle= v_\chi (1,\om,\ob)\,,\quad \langle\chi^*\rangle= v_\chi (1,\ob,\om)\,,
\ee
couple with $\tau_R$ and $\mu_R$ respectively. Using Eqs.~(\ref{eq:chichis}, \ref{eq:chivevs}), we obtain
\be\label{eq:chivevc}
\edit \langle(\chi^*\chi)_{\rtp}\rangle= -v_\chi^2 (1,1,1)\,,
\ee
which couples with $e_R$. Substituting the VEVs, Eqs.~(\ref{eq:chivevs}, \ref{eq:chivevc}) and the Higgs VEV, Eq.~(\ref{eq:vevh}),  in Eq.~(\ref{eq:masstermc}), we obtain the charged-lepton mass term after SSB,
\be\label{eq:leptcontrib}
\bar{l}_LM_l l_R,
\ee
where 
\be\label{eq:leftright}
l_L =(e_L, \mu_L, \tau_L)^T,\quad l_R=(e_R, \mu_R, \tau_R)^T,
\ee
and
\be\label{eq:mc}
\edit M_l =v \frac{v_\chi}{\Lambda}\left(\begin{matrix} -y_e \frac{v_\rho v_\chi}{\Lambda^2} & y_\mu \frac{v_\rho}{\Lambda} & y_\tau\\
	-y_e \frac{v_\rho v_\chi}{\Lambda^2} & \ob y_\mu \frac{v_\rho}{\Lambda} & \om y_\tau\\
	-y_e \frac{v_\rho v_\chi}{\Lambda^2} & \om y_\mu \frac{v_\rho}{\Lambda} & \ob y_\tau
\end{matrix}\right).
\ee
is the charged-lepton mass matrix.

\subsection{The neutrino mass matrices}
\label{subsec:neutrino}
The Dirac mass term for the neutrinos is $\edit y_\nu \bar{L} \nr \frac{\rho}{\Lambda}\widetilde{H}$. Substituting the VEVs of $H$ and $\edit \rho$ in this term, we obtain 
\be\label{eq:dircontrib}
\bar{\nu_L} M_D \nr,
\ee
where 
\be\label{eq:leftrightneutrino}
\nu_L =(\nu_e, \nu_\mu, \nu_\tau)^T,\quad N=(N_1, N_2, N_3)^T,
\ee
and
\be\label{eq:mdmat}
\edit M_D= v  \frac{v_\rho}{\Lambda}y_\nu I.
\ee
$M_D$ is the Dirac mass matrix for the neutrinos, and is proportional to the identity, $I$, at the lowest order.

In the Majorana mass terms,
\be
\edit y_s \left( \bar{\nr^c} \nr \right)_{\rs}  s + y_\phi \left( \bar{\nr^c} \nr \right)_{\rtp} \phi,
\ee
we substitute the VEVs, Eqs.~(\ref{eq:vevs}, \ref{eq:vevphi}), to obtain
\be
\bar{N}^c M_M N
\ee
where
\be\label{eq:majornaform}
\edit M_M = v_s y_s I + v_\phi y_\phi \off
\ee
is the Majorana mass matrix with $\off$ being the off-diagonal matrix,
\be\label{eq:off}
\off=\left(\begin{matrix}
	0 & \frac{1}{2\sqrt{2}} & - \sqrt{3}\\
	\frac{1}{2\sqrt{2}} & 0 & -\frac{1}{2\sqrt{2}}\\
	- \sqrt{3} & -\frac{1}{2\sqrt{2}} & 0
\end{matrix}\right).
\ee
The effective seesaw mass matrix is given by
\be\label{eq:mss}
M_{ss}=-M_D M_M^{-1} M_D^T.
\ee
Using Eqs.~(\ref{eq:mdmat}, \ref{eq:mss}), we obtain
\be\label{eq:effseesaw}
\edit M_{ss}=-v^2 \frac{v_\rho^2}{\Lambda^2}y_\nu^2 M_M^{-1}.
\ee


\section{Phenomenology}
\label{sec:phen}
We diagonalise the charged-lepton mass matrix, $M_l$, Eq.~(\ref{eq:mc}), using the $3\times 3$-trimaximal matrix, $U_\omega$, Eq.~(\ref{eq:tm}), 
\be\label{eq:chargeddiag}
U_\omega \,M_l\text{diag}\,(-1, 1, 1) = \text{diag}(m_e, m_\mu, m_\tau),
\ee
where $m_e= \sqrt{3}y_e v \frac{v_\chi^2v_\rho}{\Lambda^3}$, $m_\mu= \sqrt{3}y_\mu v \frac{v_\chi v_\rho}{\Lambda^2}$ and $m_\tau= \sqrt{3}y_\tau v \frac{v_\chi}{\Lambda}$ are the charged-lepton masses. {\edit Let us assume that the expansion parameters $\frac{v_\chi}{\Lambda}$ and $\frac{v_\rho}{\Lambda}$ are approximately equal to $0.012$. With $m_e\approx0.5~\text{MeV}$, $m_\mu \approx 0.1~\text{GeV}$, $m_\tau\approx1.8~\text{GeV}$ and $v\approx 170~\text{GeV}$, we obtain the Yukawa-like couplings to be of the order of one, $y_e\approx 1.0$, $y_\mu\approx 2.4$ and $y_\tau\approx 0.5$. We assume that the couplings in the neutrino sector $y_\nu$, $y_s$ and $y_\phi$ are also of the order of one.}

To diagonalise the effective seesaw mass matrix, $M_{ss}$, we study the diagonalisation of $\Theta$, Eq.~(\ref{eq:off}). Applying the $(13)$-bimaximal matrix, 
\be
\Uac=\left(\begin{matrix}
	\frac{1}{\sqrt{2}} & 0 & \frac{-1}{\sqrt{2}}\\[3pt]
	0 & 1 & 0\\[3pt]
	\frac{1}{\sqrt{2}} & 0 & \frac{1}{\sqrt{2}}
\end{matrix}\right),
\ee
on $\Theta$, we obtain,
\begin{align}
\begin{split}
\Uac^T  \Theta \, \Uac &=\left(\begin{matrix}
-\sqrt{3} & 0 & 0\\[3pt]
0 & 0 & -\frac{1}{2}\\[3pt]
0 & -\frac{1}{2} & \sqrt{3}
\end{matrix}\right)\\[2pt]
&=\frac{\sqrt{3}}{2} I + \left(\begin{matrix}
-\frac{3\sqrt{3}}{2} & 0 & 0\\[3pt]
0 & -\cos \alpha & \sin \alpha\\[3pt]
0 & \sin \alpha & \cos \alpha
\end{matrix}\right),
\end{split}
\end{align}
where $\alpha = -\frac{\pi}{6}$. The above matrix is diagonalised as follows:
\be\label{eq:thetadiag}
\Ubc^T \Uac^T  \Theta \, \Uac \Ubc = \text{diag}\left(-\sqrt{3}, \frac{\sqrt{3}}{2}-1, \frac{\sqrt{3}}{2}+1 \right),
\ee
where
\be
\Ubc=\left(\begin{matrix}
	1 & 0 & 0\\
	0 & \cos \theta & \sin \theta\\
	0 & -\sin  \theta & \cos \theta
\end{matrix}\right) \quad \text{with} \quad \theta = -\frac{\pi}{12}.
\ee

With the help of the above result, we diagonalise the effective seesaw mass matrix, Eq.~(\ref{eq:effseesaw}),
\begin{align}\label{eq:finaldiag}
\begin{split}
\Ubc^T \Uac^T M_{ss} \Uac \Ubc &= {\edit -v^2\frac{v_\rho^2}{\Lambda^2}y_\nu^2 \Ubc^T \Uac^T M_M^{-1}\Uac \Ubc  }\\
&= \text{diag}\left(m_1, m_2, m_3\right),
\end{split}
\end{align}
where $m_1$, $m_2$, $m_3$ are the light neutrino masses,
\begin{align}
m_1 &= \left(\sqrt{3}-r\right)^{-1}m, \label{eq:m1}\\
m_2 &= \left(-\frac{\sqrt{3}}{2}+1-r \right)^{-1}m, \label{eq:m2}\\
m_3 &= \left(-\frac{\sqrt{3}}{2}-1-r\right)^{-1}m \label{eq:m3},
\end{align}
with 
\be\label{eq:param}
\edit r=\frac{v_sy_s}{v_\phi y_\phi}, \quad m= \frac{v^2 v_\rho^2 \,y_\nu^2}{v_\phi \Lambda^2 \,y_\phi}.
\ee
To obtain Eq.~(\ref{eq:finaldiag}) from Eqs.~(\ref{eq:majornaform}, \ref{eq:effseesaw}, \ref{eq:thetadiag}), we have used the fact that a real symmetric matrix, as well as its inverse, are diagonalised by the same orthogonal matrix. The eigenvalues of the inverse matrix are simply the inverse of the eigenvalues of the original matrix.

The PMNS matrix is obtained by multiplying the diagonalising matrices of the charged-lepton mass matrix, Eq.~(\ref{eq:chargeddiag}), and the effective seesaw mass matrix, Eq.~(\ref{eq:finaldiag}),
\be\label{eq:pmnsformula}
U= U_\omega \Uac \Ubc.
\ee
The tribimaximal matrix, Eq.~(\ref{eq:tbm}), can be obtained in terms of $U_\omega$ and $\Uac$,
\be\label{eq:tbmformula}
U_\omega \Uac= \text{diag}(1, \om, \ob) U_{TBM} \text{diag}(1, 1, i).
\ee
Substituting Eq.~(\ref{eq:tbmformula}) in Eq.~(\ref{eq:pmnsformula}), we obtain
\be\label{eq:i}
U= \text{diag}(1, \om, \ob) U_{TBM} \text{diag}(1, 1, i) \Ubc
\ee
Comparing the above equation with Eq.~(\ref{eq:tma}), we get
\be\label{eq:upmns}
U= \text{diag}(1, \om, \ob) \, \Utm \,\text{diag}(1, 1, i).
\ee
Note that a $\text{TM}_1$ mixing scheme with $\theta=\frac{\pi}{12}$ and $\zeta=\frac{\pi}{2}$ was proposed in Ref.~\cite{Zhou:2012zj}. The unobservable phases, $\text{diag}(1, \om, \ob)$, can be ignored. The Majorana phases, $\text{diag}(1, 1, i)$, are potentially observable in the neutrinoless double-beta decay experiments. 

Substituting the values of $\theta=\frac{-\pi}{12}$ and $\zeta=\frac{\pi}{2}$ in Eqs.~(\ref{eq:theta13}, \ref{eq:theta12}, \ref{eq:theta23}, \ref{eq:jcp}), we obtain
\begin{align}
\sin^2 \theta_{13}& = \frac{1}{3}\sin^2 \frac{-\pi}{12}=0.0223\,,\\
\sin^2 \theta_{12}& = 1-\frac{2}{3-\sin^2\frac{-\pi}{12}}=0.318\,,\\
\sin^2 \theta_{23}& = \frac{1}{2}\,, \label{eq:tpws}\\
J& = -\frac{1}{12 \sqrt{6}}.
\end{align}
Comparing the values of $J$ and the mixing angles with Eq.~(\ref{eq:deltacp}), we obtain
\be
\delta_{\text{CP}}=-\frac{\pi}{2}.
\ee
{\edit The VEV of the flavon $\chi$, Eq.~(\ref{eq:vevchi}), spontaneously breaks CP symmetry and produces the complex charged-lepton mass matrix, Eq.~(\ref{eq:mc}). Since the neutrino mass matrices are real, the charge-lepton sector becomes the sole source of CP violation in the model. $U_\omega$, which diagonalises the charged-lepton mass matrix, results in the appearance of $i$ in Eq.~(\ref{eq:i}), which manifests as $\zeta=\frac{\pi}{2}$ in $\text{TM}_1$ mixing. This leads to $\delta_\text{CP}=-\frac{\pi}{2}$.} From Eq.~(\ref{eq:tma}), it is clear that $\zeta=\pm\frac{\pi}{2}$ leads to the $\mu$ and $\tau$ rows of the mixing matrix being conjugate to each other. This property, termed as the $\mu\text{-}\tau$ reflection symmetry~\cite{Harrison:2002kp,Harrison:2002et,Ma:2002ce,Babu:2002dz,Grimus:2003yn}, results in maximal atmospheric mixing, i.e.~$\sin^2 \theta_{23}= \frac{1}{2}$ as predicted by our model. {\edit The origin of $\mu\text{-}\tau$ reflection symmetry becomes apparent when we express the neutrino effective seesaw mass matrix $M_{ss}$, Eq.~(\ref{eq:effseesaw}), in the basis where the charged-lepton mass matrix is diagonal, i.e.~$U_\omega^*M_{ss}U_\omega^\dagger$. This matrix remains invariant under the $C_2$ symmetry generated by complex conjugation and simultaneous exchange of $\mu$-$\tau$ rows and columns. This $C_2$ symmetry manifests as the $\mu\text{-}\tau$ reflection symmetry of the mixing matrix.}

Our predictions, $\sin^2 \theta_{13} =0.0223$ and $\sin^2 \theta_{12}=0.318$, are within the $1\sigma$ ranges of the global fit, Eqs.~(\ref{eq:gf13}, \ref{eq:gf12}). The prediction $\sin^2 \theta_{23}= \frac{1}{2}$ is expected in any mixing scenario with $\mu$-$\tau$ reflection symmetry. There are indications that the atmospheric mixing may not be maximal, which gives rise to the problem of the octant degeneracy. However, at the $3\sigma$ level, the allowed range is $0.428<\sin^2 \theta_{23}<0.624$. Therefore, $\sin^2 \theta_{23} =\frac{1}{2}$ is not ruled out. CP violation in the lepton sector is still an open problem and we do not have an accurate measurement of $\delta_{\text{CP}}$. At the $3\sigma$ level, we have a large range, $135^\circ<\delta_{\text{CP}}<366^\circ$, and the maximal value of $\delta_{\text{CP}}=-\frac{\pi}{2}$ lies within this range.

\begin{figure}
	\includegraphics[width=65mm]{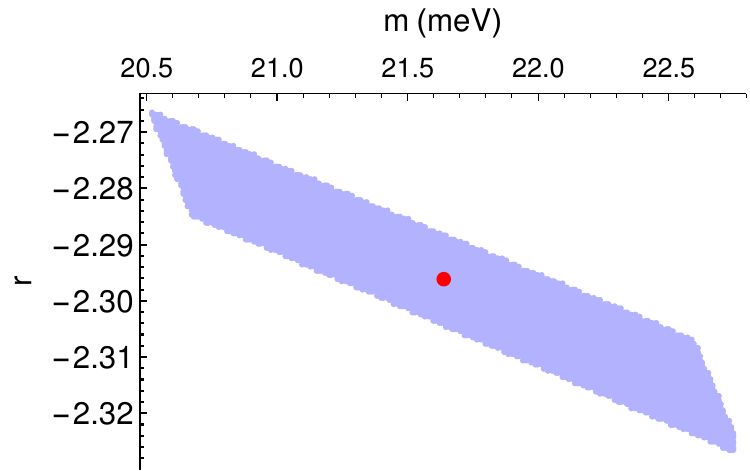}
	\caption{The allowed ranges of the parameters $r$ and $m$ at the $3\sigma$ level. The red point corresponds to the best fit.}
	\label{fig:params}
\end{figure}

\begin{figure}
	\includegraphics[width=65mm]{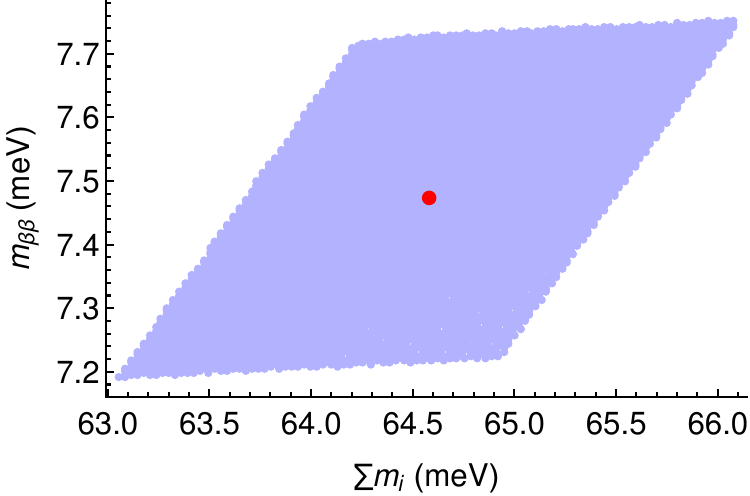} 
	\caption{$\Sigma m_i$ and $m_{\beta\beta}$ as predicted by the model}
	\label{fig:masses}
\end{figure}

The neutrino masses, Eqs.~(\ref{eq:m1}, \ref{eq:m2}, \ref{eq:m3}) are given in terms of two free parameters, $r$ and $m$. Using the known mass-squared differences,
\begin{align}
\Delta m^2_{21}&=67.9\rightarrow 80.1\text{ meV}^2,\\
\Delta m^2_{31}&=2432 \rightarrow 2618\text{ meV}^2,
\end{align}
we can fit these parameters and predict the individual light neutrino masses. The allowed parameter space is shown in Figure~\ref{fig:params}. The resulting values of the individual neutrino masses are
\begin{align}
m_1&=5.1\rightarrow5.6\text{ meV},\label{eq:m1val}\\
m_2&=8.5\rightarrow9.3\text{ meV},\label{eq:m2val}\\
m_3&=49.4\rightarrow51.3\text{ meV},\label{eq:m3val}.
\end{align}
{\edit From Figure~\ref{fig:params}, Eqs.~(\ref{eq:param}) and given $\frac{v_\rho}{\Lambda} \approx 0.01$, we can estimate that the flavon VEVs $v_\phi$ and $v_s$ are of the order of $10^{11}~\text{GeV}$.}

The sum of the neutrino masses is constrained by cosmological observations. The upper bounds provided by these observations are a few hundreds of meV~\cite{Choudhury:2018byy,Loureiro:2018pdz}, the most stringent bound being $\Sigma m_i <78$~meV~\cite{Choudhury:2018byy}. From Eqs.~(\ref{eq:m1val})-(\ref{eq:m3val}), we predict
\be\label{eq:sum}
\Sigma m_i = 63.1\rightarrow66.1\text{ meV}.
\ee
This range is below the experimental bounds.

The neutrinoless double-beta decay experiments seek to determine whether the neutrinos are Majorana particles. If the neutrinoless double-beta decay is observed, the measurement of half-life leads to the determination of the effective mass,
\be\label{eq:dbetaformula}
m_{\beta\beta}=U_{e1}^2 m_1 + U_{e2}^2 m_2 +U_{e3}^2 m_3,
\ee
where $U_{e1}$, $U_{e2}$ and $U_{e3}$ are the elements of the first row of the PMNS matrix. Substituting the values of these elements from Eq.~(\ref{eq:upmns}) in Eq.~(\ref{eq:dbetaformula}), we obtain
\be
m_{\beta\beta}=\frac{2}{3} m_1 + \frac{1}{12}\left(2+\sqrt{3}\right) m_2 +\frac{1}{12}\left(2-\sqrt{3}\right) m_3.
\ee
Given our predicted masses, Eqs.~(\ref{eq:m1val}, \ref{eq:m2val}, \ref{eq:m3val}), we obtain
\be
m_{\beta\beta}=7.2\rightarrow7.7\text{ meV}.
\ee
This range is well below the upper bounds set by the recent $0\nu\beta\beta$ experiments~\cite{Anton:2019wmi,Azzolini:2019tta}. Figure~\ref{fig:masses} shows the ranges of $\Sigma m_i$ and $m_{\beta\beta}$ predicted by our model along with the best-fit point.


\section{Uniquely Defining the VEVs}
\label{sec:defvev}

\subsection{Orbits and stabilisers}
\label{subsec:ons}

\begin{figure}
	\begin{tabular}{ccc}
		\includegraphics[width=40mm,trim={2.8cm 3.5cm 1.0cm 0.0cm},clip]{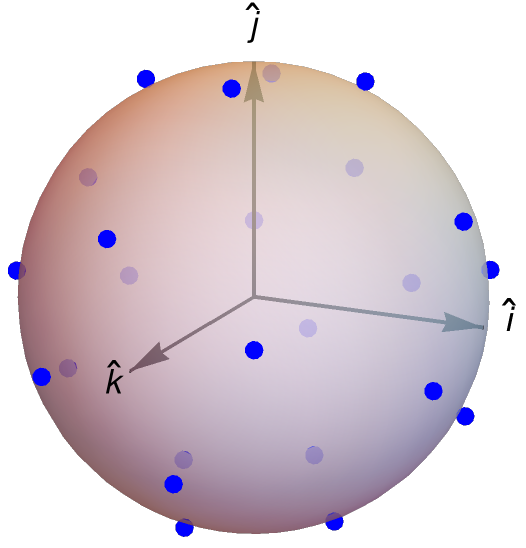}&\includegraphics[width=40mm,trim={2.8cm 3.5cm 1.0cm 0.0cm},clip]{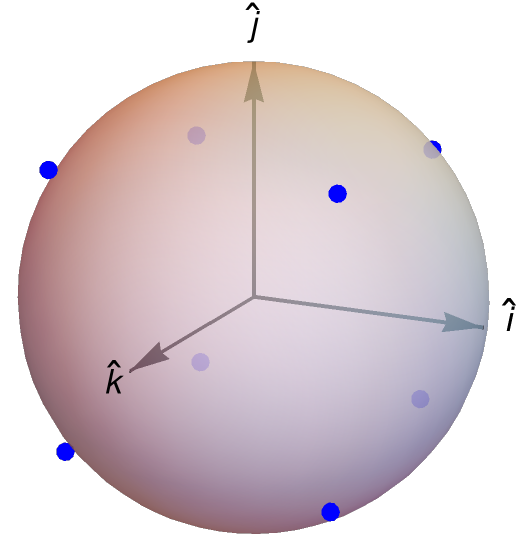} \\
		(a) & (b) \\[6pt]
		\includegraphics[width=40mm,trim={2.8cm 3.5cm 1.0cm 0.0cm},clip]{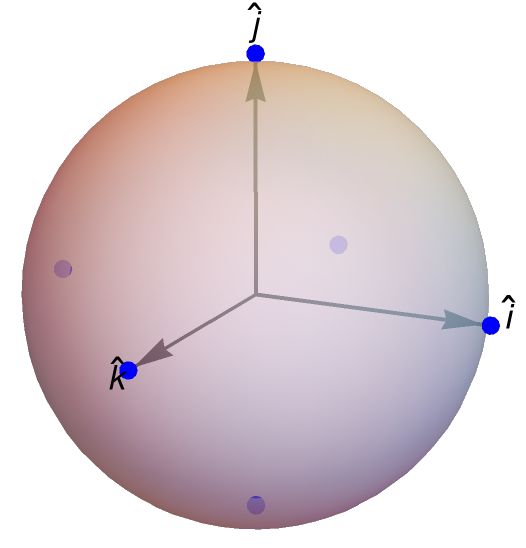}&\includegraphics[width=40mm,trim={2.8cm 3.5cm 1.0cm 0.0cm},clip]{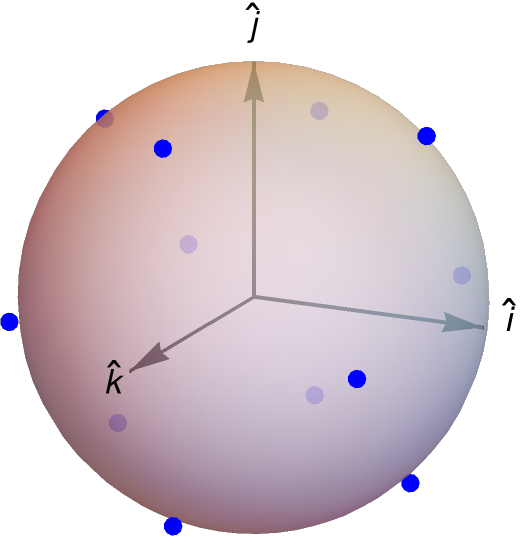}\\
		(c) & (d)  \\[6pt]
	\end{tabular}
	\caption{Orbits under $\rt$ of $S_4$}
	\label{fig:orb3}
\end{figure}

In this section, we describe the concepts of orbits and stabilisers by using the action of the triplet representations of $S_4$ as examples. We utilise these concepts later in the paper. Various representations of a group act on the corresponding vector spaces. The triplet representations of $S_4$ act on the three-dimensional real space. Consider the set of all unit vectors forming a sphere in the 3-D real space. The action of an element of $\rt$ of $S_4$ on any unit vector rotates it from one point on the sphere to another. Since $\rt$ has 24 distinct elements, the group action on a given point leads to a set of 24 points on the sphere in the most general case, as shown in Figure~\ref{fig:orb3}(a). This set of points forms the orbit of the given point. The group action on any point within the orbit produces the same orbit; therefore, the orbit remains closed under the group action.

It need not be the case that all the orbits have their cardinality equal to the number of distinct elements in the representation. Group action on $x=\frac{1}{\sqrt{3}}(1,1,1)$ will produce the points,
\begin{align}\label{eq:orb1}
\begin{split}
\frac{1}{\sqrt{3}}&\left\{(\pm1,\pm1,\pm1),(\pm1,\pm1,\mp1),\right.\\
&\left.\quad(\pm1,\mp1,\pm1),(\mp1,\pm1,\pm1)\right\},
\end{split}
\end{align}
which form the vertices of a cube, Figure~\ref{fig:orb3}(b). Rather than 24, here we have only 8 points in the orbit because each point on the cube remains invariant under the action of certain elements of the group. The set of such elements form a subgroup and it is termed as the stabiliser of the point. The stabiliser of  $\frac{1}{\sqrt{3}}(1,1,1)$ is $\{Q, Q^2, I\}$ which forms a $C_3$ subgroup of $S_4$. We have four such $C_3$ subgroups, corresponding to the four pairs of opposite vertices of the cube. The orbit-stabiliser theorem states that
\be\label{eq:ost}
|\text{Orb}(x)| = \frac{|G|}{|\text{Stab}(x)|},
\ee
where Orb$(x)$ and Stab$(x)$ denote the orbit and the stabiliser respectively of a point x and $||$ denotes the cardinality. In the case of the cube, $|\text{Stab(x)}|=3$; therefore, we have
\be
|\text{Orb}(x)| = \frac{24}{3} = 8,
\ee
which is consistent with the number of vertices of the cube.

The group action on $x=(1,0,0)^T$ produces an octahedron, 
\be\label{eq:orb2}
\left\{(\pm1,0,0), (0,\pm1,0), (0,0,\pm1)\right\},
\ee
Figure~\ref{fig:orb3}(c). The stabiliser of $x=(1,0,0)^T$ is the $C_4$ subgroup, $\{R, R^2, R^3, I\}$. Using the orbit-stabiliser theorem, we obtain the cardinality of the orbit (the number of vertices of the octahedron) to be $\frac{24}{4}=6$. 

The group action on $x=\frac{1}{\sqrt{2}}(1,0,1)$ produces a cuboctahedron, 
\begin{align}\label{eq:orb3}
\begin{split}
\frac{1}{\sqrt{2}}&\left\{(\pm1,0,\pm1),(\pm1,\pm1,0),(0,\pm1,\pm1),\right.\\
&\left. (\pm1,0,\mp1),(\pm1,\mp1,0),(0,\pm1,\mp1)\right\},
\end{split}
\end{align}
Figure~\ref{fig:orb3}(d). In this case, the stabiliser of $x$ is the $C_2$ subgroup, $\{P, I\}$ and we obtain the number of vertices of the cuboctahedron to be $\frac{24}{2}=12$. 

The cube, Eq.~(\ref{eq:orb1}), the octahedron, Eq.~(\ref{eq:orb2}), and the cuboctahedron, Eq.~(\ref{eq:orb3}), are the only `uniquely defined' orbits that can be constructed in the case of the representation $\rt$. They are unique in the sense that the orientations of their elements are fully defined by their respective residual symmetries (subgroups of $S_4$).  For any other orbit of $\rt$, the stabiliser of a point in the orbit will be the trivial group and the orbit will have the cardinality of $24$.  Such orbits do not form uniquely defined geometric objects like the cube.

\begin{figure}
	\begin{tabular}{ccc}
		\includegraphics[width=40mm,trim={2.8cm 3.5cm 1.0cm 0.0cm},clip]{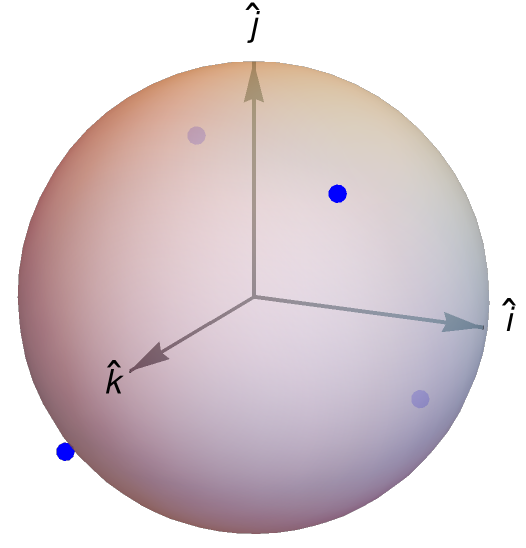}&\includegraphics[width=40mm,trim={2.8cm 3.5cm 1.0cm 0.0cm},clip]{octpnts.pdf} \\
		(a) & (b) \\[6pt]
		\includegraphics[width=40mm,trim={2.8cm 3.5cm 1.0cm 0.0cm},clip]{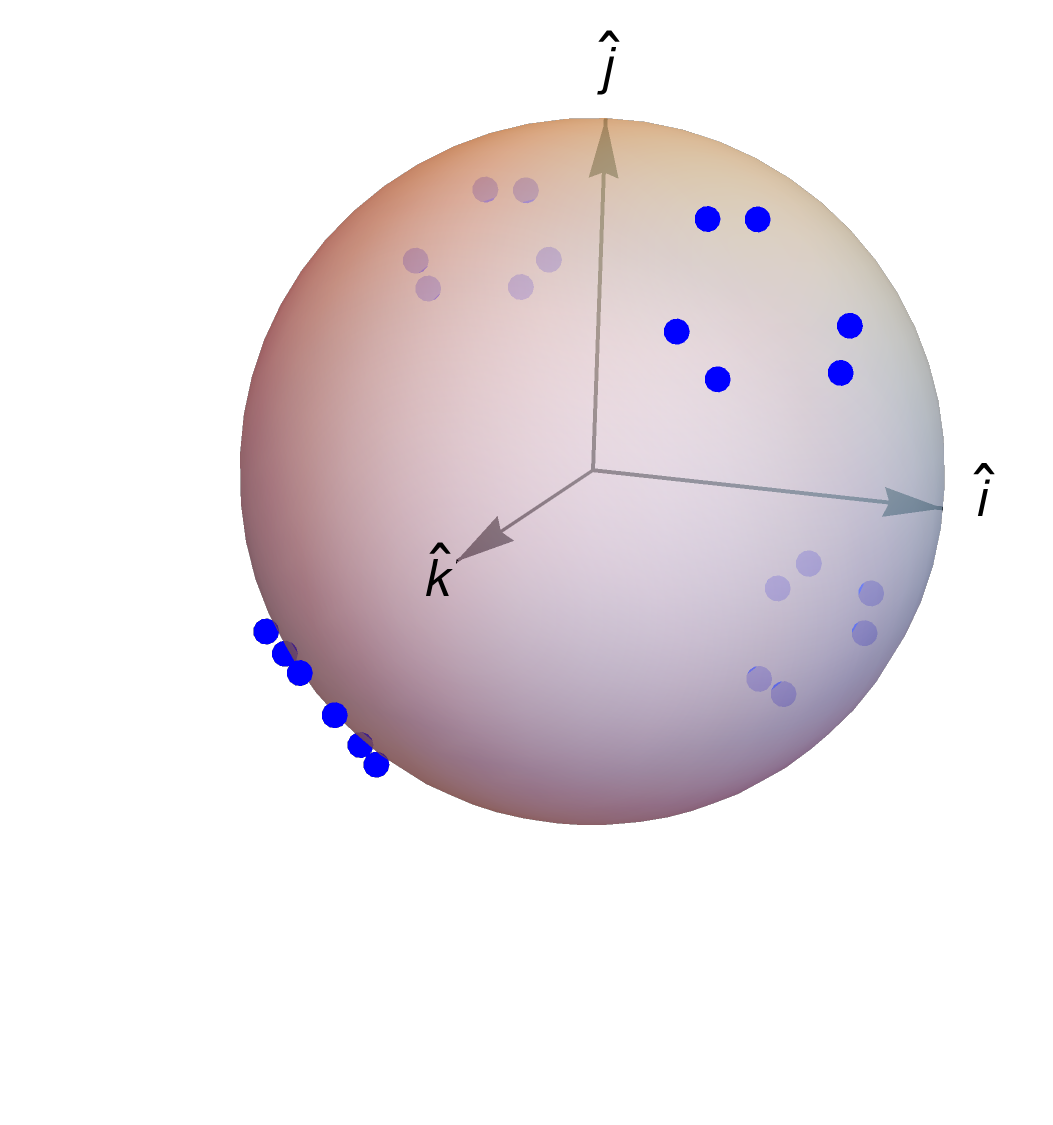}&\includegraphics[width=40mm,trim={2.8cm 3.5cm 1.0cm 0.0cm},clip]{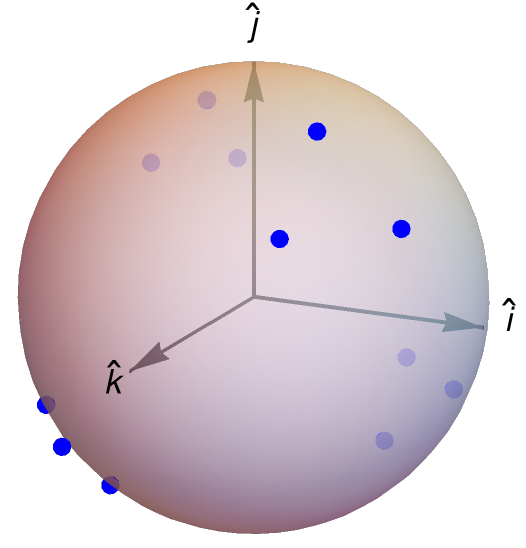}\\
		(c) & (d)  \\[6pt]
	\end{tabular}
	\caption{Orbits under $\rtp$ of $S_4$}
	\label{fig:orb3p}
\end{figure}

Now let us analyse the representation $\rtp$. We have $P(\rtp)=-P, Q(\rtp)=Q, R(\rtp)=-R$, Eqs.~(\ref{eq:tripletpgens}). This representation is the symmetry group of a tetrahedron which includes 12 proper rotations and 12 improper rotations. The action of $\rtp$ on the point $\frac{1}{\sqrt{3}}(1,1,1)$ produces the orbit,
\be\label{eq:orb4a}
\frac{1}{\sqrt{3}}\left\{(1,1,1),(1,-1,-1),(-1,1,-1),(-1,-1,1)\right\},
\ee
which forms the vertices of the tetrahedron, Figure~\ref{fig:orb3p}(a). The point $\frac{1}{\sqrt{3}}(1,1,1)$ has the stabiliser $\{ Q, Q^2, -QPQ^2R^2$, $-QPQ^2R^2Q, -QPQ^2R^2Q^2, I \}$ which forms the subgroup $D_6$ (the dihedral group with 6 elements). The cardinality of the orbit (the number of vertices of the tetrahedron) is $\frac{24}{6}=4$. The action of $\rtp$ on the point $\frac{1}{\sqrt{3}}(-1,-1,-1)$ produces the orbit,
\be\label{eq:orb4b}
\frac{1}{\sqrt{3}}\left\{(-1,-1,-1),(-1,1,1),(1,-1,1),(1,1,-1)\right\},
\ee
forming a second tetrahedron which is the space inversion of the first\footnote{Combining the two tetrahedra, we obtain a cube.}.

The group action of $\rtp$ on the point $(1,0,0)$ produces the orbit,
\be\label{eq:orb5}
\left\{(\pm1,0,0), (0,\pm1,0), (0,0,\pm1)\right\},
\ee
which forms an octahedron, Figure~\ref{fig:orb3p}(b). The stabiliser of the point $(1,0,0)^T$ is $\{R^2,-QPQ^2, -QPQ^2R^2, I\}$ which forms the subgroup $C_2\times C_2$. The cardinality of the orbit is $\frac{24}{2\times2}=6$. The tetrahedra, Eqs.(\ref{eq:orb4a},\ref{eq:orb4b}) and the octahedron, Eq.~(\ref{eq:orb5}) are the only unique orbits that can be constructed in the case of the representation $\rtp$.

The action of $\rtp$ on a random point produces an orbit with $24$ elements. This situation is shown in Figure~\ref{fig:orb3p}(c). Here we have a trivial stabiliser. For $\rtp$, a point having a non-trivial stabiliser is not the sufficient condition for its orbit to be uniquely defined. Consider a point $ \propto \left(\frac{1}{\sqrt{2}}\sin \alpha,-2\cos \alpha,-\frac{1}{\sqrt{2}}\sin \alpha\right)$, Eq.~(\ref{eq:vevalpha}). This point remains invariant under the action of $\{-P, I\}$ which is a non-trivial stabiliser. The orbit of this point has $\frac{24}{2}=12$ elements and is shown in Figure~\ref{fig:orb3p}(d). It is clear that this orbit is not unique, rather it depends on the arbitrary parameter $\alpha$.

Let us study the VEVs used in our model in terms of their residual symmetries and check whether these VEVs can be uniquely defined. Consider the VEV of the doublet flavon, $\langle \eta \rangle \propto \left(-\frac{1}{2}, \frac{\sqrt{3}}{2}\right) $, Eq.~(\ref{eq:veveta}). $P(\rd)$, Eq.~(\ref{eq:doubletgens}), generates the $C_2$ residual symmetry of $\langle \eta \rangle$,
\be
P(\rd) \langle \eta \rangle = \langle \eta \rangle.
\ee
This residual symmetry uniquely defines $\langle \eta \rangle$\footnote{\label{note1}up to a proportionality constant.}.

The flavon $\chi$ transforms as $\rt$ under $S_4$. It is also assigned a $C_3$ symmetry ($\om$), Table~\ref{tab:flavourcontent}, and hence it has complex degrees of freedom. Its VEV, $\langle \chi \rangle \propto \left(1, \om, \ob\right) $, Eq.~(\ref{eq:vevchi}), remains invariant under the action of $\om Q^2$,
\be
\om Q^2 \langle \chi \rangle = \langle \chi \rangle.
\ee
The group element $\om Q^2$ generates the $C_3$ subgroup $\{\om Q^2, \ob Q, I\}$, which forms the residual symmetry of $\langle \chi \rangle$. This $C_3$ residual symmetry uniquely defines $\langle \chi \rangle$\textsuperscript{\ref{note1}}.

The flavon $\phi$ transforms as $\rtp$ under $S_4$. Its VEV $\langle \phi \rangle \propto \left(-\frac{1}{2\sqrt{2}}, -\sqrt{3}, \frac{1}{2\sqrt{2}}\right) $, Eq.~(\ref{eq:vevphi}), has the following $C_2$ residual symmetry:
\be
P(\rtp) \langle \phi \rangle = -P \langle \phi \rangle = \langle \phi \rangle.
\ee
This symmetry ensures that the third component of the VEV is negative of the first component. However, it does not determine the value of the second component in relation to the others. In fact, this value can not be determined by any of the symmetries of $S_4$. Rather we obtained it in terms of $\alpha$, Eq.~(\ref{eq:vevalpha}), which is a continuous function of the parameters, Eq.~(\ref{eq:alphaf}), appearing in the flavon potential, Eq.~(\ref{eq:phipot}). This potential results in a set of minima that forms the orbit shown in Figure~\ref{fig:orb3p}(d). We tuned the parameters in the potential to make $\alpha$ equal to $-\frac{\pi}{6}$. This tuning of the parameters changes the orbit (resizes the four small triangles in Figure~\ref{fig:orb3p}(d)) as a continuous function and enables us to choose the VEV from among an infinite set of possible alignments. We argue that such a dependence of the VEV on the parameters of the potential goes against the spirit of using discrete symmetries in model building. We can resolve this problem by incorporating additional symmetries in the model. A way to achieve this is by utilising the recently proposed framework of the auxiliary group~\cite{Krishnan:2019ftw}.

\subsection{The Framework of the Auxiliary Group}
\label{sec:xxx}

Various discrete subgroups of $U(3)$ with an irreducible triplet representation ($A_4$ and the higher members of the $\Delta(3n^2)$ series, $S_4$ and the higher members of the $\Delta(6n^2)$ series, $\Sigma(72\times 3)$, $PSL(2,7)$, etc.) combined with Abelian discrete groups ($C_2$, $C_3$, etc.) have been studied in the literature as discrete flavour groups. In those studies, the subgroups of $U(3)$ with triplets are considered because the fermions exist in three families. However, it was shown in Refs.~\cite{Babu:2010bx,Holthausen:2011vd,Holthausen:2012wz} that by going beyond the $U(3)$-subgroup paradigm we can effectively avoid certain undesirable cross terms in the flavon potential. These papers used an enlarged flavor group constructed as a semidirect product in which the conventional flavor group (the direct product of a subgroup of $U(3)$ with Abelian groups) appears as the quotient. In Ref.~\cite{Krishnan:2019ftw,2011.11653}, a simpler construction, in which the semidirect product was replaced with a direct product, was studied. We named this construction the `framework of the auxiliary group' and used the notation 
\be
G_f = G_r \times G_x.
\ee
where $G_f$ is the enlarged flavor group, $G_r$ is the conventional flavor group and $G_x$ is the so-called auxiliary group. The auxiliary group ($G_x$) is defined as the group under which the fermions remain invariant. Only the flavons transform nontrivially under it. On the other hand, both the fermions and the flavons transform nontrivially under the `conventional' flavour group ($G_r$). In the flavon sector, to obtain VEVs similar to Eq.~(\ref{eq:vevphi}), we incorporate symmetries in addition to those originating from $G_r$ with the help of $G_x$. Since $G_x$ does not act on the fermions, it need not be a subgroup of $U(3)$. 

An effective multiplet of a given irreducible representation of $G_r$ can be obtained by coupling together several elementary flavons transforming as irreducible multiplets under $G_r \times G_x$. The vacuum alignments of these elementary multiplets are constructed in such a way that they are invariants under the action of various subgroups of $G_r \times G_x$, i.e.~these subgroups form the residual symmetries of the VEVs of the respective multiplets. Moreover, these VEVs are uniquely defined by their residual symmetries. Even though the VEV of the effective multiplet of $G_r$ may not have any residual symmetry under $G_r$, it will be uniquely defined in terms of the residual symmetries of its constituent elementary multiplets (transforming under $G_r \times G_x$). We will get a clearer understanding of this framework over the course of the following sections when we apply it to our model.

\section{The Group $Y_{24}$}
\label{sec:aux}

In this section, we construct a group that we call $Y_{24}$ and briefly study its structure. $Y_{24}$ is later used to recast our model in the framework of the auxiliary group. 

Consider the group generated by
\be\label{eq:ygens}
A=\left(\begin{matrix}
	0 & 0 & 0 & \tb & 0 & 0\\
	0 & 0 & 0 & 0 & i \ob & 0\\
	0 & 0 & 0 & 0 & 0 & -i \om\\
	\ta & 0 & 0 & 0 & 0 & 0\\
	0 & 1 & 0 & 0 & 0 & 0\\
	0 & 0 & 1 & 0 & 0 & 0
\end{matrix}\right), \,\,B=
\left(\begin{matrix}
	0 & \tb & 0 & 0 & 0 & 0\\
	0 & 0 & 1 & 0 & 0 & 0\\
	1 & 0 & 0 & 0 & 0 & 0\\
	0 & 0 & 0 & 0 & \ta & 0\\
	0 & 0 & 0 & 0 & 0 & -i \om\\
	0 & 0 & 0 & i \ob & 0 & 0
\end{matrix}\right),
\ee
where $\ta=e^{i\frac{\pi}{4}}$ and $\tb=e^{-i\frac{\pi}{4}}$ are the complex eighth roots of unity. Using  $A$ and $B$, we obtain the group elements $D=(AB)^3$ and $E= (AB)^2$ which are given by
\begin{align}
D=\left(\begin{matrix}
0 & 0 & 0 & 1 & 0 & 0\\
0 & 0 & 0 & 0 & 1 & 0\\
0 & 0 & 0 & 0 & 0 & 1\\
1 & 0 & 0 & 0 & 0 & 0\\
0 & 1 & 0 & 0 & 0 & 0\\
0 & 0 & 1 & 0 & 0 & 0
\end{matrix}\right), \quad E=
\left(\begin{matrix}
0 & 0 & 1 & 0 & 0 & 0\\
1 & 0 & 0 & 0 & 0 & 0\\
0 & 1 & 0 & 0 & 0 & 0\\
0 & 0 & 0 & 0 & 0 & 1\\
0 & 0 & 0 & 1 & 0 & 0\\
0 & 0 & 0 & 0 & 1 & 0
\end{matrix}\right).
\end{align}
We also obtain the following diagonal group elements consisting of complex phases,
\begin{align}
\Cwa&= (DA)^8=  \text{diag}(1,1,1,1,\om,\ob),\\
\Cwb&= E(DA)^8E^2=  \text{diag}(1,1,1,\ob,1,\om),\\
\Cwc&= D\Cwa D=  \text{diag}(1,\om,\ob,1,1,1),\\
\Cwd&= D\Cwb D=  \text{diag}(\ob,1,\om,1,1,1),\\
\Cta&= (DA)^9=  \text{diag}(\ta,1,1,\tb,i,-i),\\
\Ctb&= E(DA)^9E^2=  \text{diag}(1,\ta,1,-i,\tb,i),\\
\Ctc&= E^2(DA)^9E=  \text{diag}(1,1,\ta,i,-i,\tb).
\end{align}

We obtain the structure of the group generated by $A$ and $B$ with the help of the above-mentioned group elements, $D$, $E$ and $C_i$. We have
\be\label{eq:decommute}
D^2 = I, \quad E^3 = I, \quad DE = ED.
\ee
D and E form the cyclic groups $C_2$ and $C_3$ respectively and they commute with each other. Therefore, the two of them form the group $C_6$. The diagonal group elements $\Cwa$, $\Cwb$, $\Cwc$ and $\Cwd$ generate four separate $C_3$ groups. Similarly, the elements $\Cta$, $\Ctb$ and $\Ctc$ generate three separate $C_8$ groups. Multiplication of $D$ and $E$ with the diagonal group elements leads to
\begin{align}\label{eq:dediag}
\begin{split}
&D \Cwa = \Cwc D, \\
&D \Cwb = \Cwd D, \\
&D \Cwc = \Cwa D, \\
&D \Cwd= \Cwb D, \\
&D \Cta = \Cta^7\Ctb^2\Ctc^6 D, \\
&D \Ctb = \Cta^6\Ctb^7\Ctc^2 D, \\
&D \Ctc = \Cta^2\Ctb^6\Ctc^7 D, 
\end{split}
\begin{split}
&E \Cwa = \Cwb E, \\
&E \Cwb = \Cwa^2\Cwb^2 E, \\
&E \Cwc = \Cwd E, \\
&E \Cwd= \Cwc^2\Cwd^2 E, \\
&E \Cta = \Ctb E, \\
&E \Ctb = \Ctc E, \\
&E \Ctc = \Cta E.
\end{split}
\end{align}
Also, the diagonal group elements, $C_i$, commute with each other,
\be\label{eq:diagcommute}
C_i C_j = C_j C_i
\ee
Eqs.~(\ref{eq:decommute}, \ref{eq:dediag}, \ref{eq:diagcommute}) prove that the group generated by $D$, $E$ and $C_i$ is finite. The generators $A$ and $B$ can be expressed in terms of $D$, $E$ and $C_i$,
\be
A=\Cwc^2\Cta^{15}\Ctb^2\Ctc^6D, \quad B= \Cwa \Cta^{15}E^2.
\ee
Since $A$ and $B$ can be expressed in terms of $D$, $E$ and $C_i$ and vice versa, the group generated by $A$ and $B$ is the same as the one generated by $D$, $E$ and $C_i$. The diagonal elements $\Cwa$, $\Cwb$, $\Cwc$, $\Cwd$, $\Cta$, $\Ctb$, $\Ctc$ form the group $C_3 \times C_3\times C_3 \times C_3\times C_8\times C_8\times C_8$ which is equivalent to $C_3 \times C_{24} \times C_{24} \times C_{24}$. As mentioned earlier, $D$ and $E$ form the group $C_6$.  Therefore, the group generated by $D$, $E$ and $C_i$ (or the one generated by $A$ and $B$) is simply the semidirect product,
\be
G(A,B)=(C_3 \times C_{24} \times C_{24} \times C_{24})\rtimes C_6.
\ee

We enlarge this group by adding another generator,
\be\label{eq:ygens1}
\Cma=\text{diag}(1,1,1,1,-1,-1).
\ee
Using $\Cma$, we obtain the group element,
\be
\Cmb=E\Cma E^2 = \text{diag}(1,1,1,-1,1,-1).
\ee
The finiteness of the enlarged group is ensured by the following equations:
\begin{align}
\begin{split}
&D \Cma = \Cma\Ctb^4\Ctc^4D, \\
&D \Cmb = \Cmb\Cta^4\Ctc^4D,
\end{split}
\begin{split}
&E \Cma = \Cmb E, \\
&E \Cmb = \Cma \Cmb E.
\end{split}
\end{align}
$\Cma$ and $\Cmb$ generate two $C_2$ groups. As a result, we obtain
\be
G(A,B, \Cma)=(C_2 \times C_2 \times C_3 \times C_{24} \times C_{24} \times C_{24})\rtimes C_6.
\ee
For convenience, we name $G(A,B, \Cma)$  as $Y_{24}$,
\be
Y_{24}=G(A,B, \Cma)=(C_2 \times C_6 \times C_{24} \times C_{24} \times C_{24})\rtimes C_6.
\ee
Any element of $Y_{24}$ can be uniquely expressed as,
\be
g_i = \Cma^{i_1}\Cmb^{i_2} \Cwa^{j_1}\Cwb^{j_2}\Cwc^{j_3}\Cwd^{j_4}\Cta^{k_1}\Ctb^{k_2}\Ctc^{k_3}D^{i_3}E^{j_5},
\ee
where $i_x=1,2$, $j_x=1,2,3$ and $k_x=1,2...8$. This group has a total of $2\times 6\times 24^3\times 6 = 995328$ elements. We used the computational package GAP~\cite{Gap4} to obtain the group with $A$, $B$ and $C_1$ as the generators. According to GAP, such a group has $995328$ elements confirming our analysis.

We name the sextet representation of $Y_{24}$ defined by the generators Eqs.~(\ref{eq:ygens},  \ref{eq:ygens1}) as $\rx$. Its tensor product with itself leads to the following expansion,
\be\label{eq:sixsix}
\rx \times \rx = \rx_0 + \rx_1 + \rx_2 + \rt + \rx_1 +\rx_2 +\rtp.
\ee
The first four multiplets in the RHS of the above equation form the symmetric part and the next three form the antisymmetric part of the tensor product. Besides $\rx$, we also utilise the representation $\rx_1$ in this paper. $\rx_1$ constructed from the symmetric part of Eq.~(\ref{eq:sixsix}) is given by
\be\label{eq:61}
\begin{split}
	&\rx_1 \equiv\\
	&\quad \left(\{a_2,b_3\}, \{a_3,b_1\}, \{a_1,b_2\},\{a_5,b_6\},\{a_6,b_4\},\{a_4,b_5\}\right)^T\\ 
\end{split}
\ee
where $a_i$ and $b_i$ form sextets transforming under $\rx$ and $\{a_i,b_j\}=a_i b_j+a_j b_i$. In the basis, Eq.~(\ref{eq:61}), the generators of $\rx_1$ are
\begin{align}\label{eq:gensixone}
\begin{split}
&A(\rx_1)\!=\!\left(\begin{matrix}
0 & 0 & 0 & 1 & 0 & 0\\
0 & 0 & 0 & 0 & -\ta \om & 0\\
0 & 0 & 0 & 0 & 0 & \ta \ob\\
1 & 0 & 0 & 0 & 0 & 0\\
0 & \ta & 0 & 0 & 0 & 0\\
0 & 0 & \ta & 0 & 0 & 0
\end{matrix}\right)\!\!, B(\rx_1)\!=\!
\left(\begin{matrix}
0 & 1 & 0 & 0 & 0 & 0\\
0 & 0 & \tb & 0 & 0 & 0\\
\tb & 0 & 0 & 0 & 0 & 0\\
0 & 0 & 0 & 0 & 1 & 0\\
0 & 0 & 0 & 0 & 0 & -\tb \ob\\
0 & 0 & 0 & \tb \om & 0 & 0
\end{matrix}\right)\!,\\
&\Cma(\rx_1)=\text{diag}(1,1,1,1,-1,-1).
\end{split}
\end{align}
This completes our discussion of the structure of the group $Y_{24}$.

\section{The Model recast using $(S_4\times C_4 \times C_3\, {\edit \times \,C_2})\times (Y_{24}\rtimes C_2\times C_2 \times C_2)$}
\label{sec:recast}

{\renewcommand{\arraystretch}{1.4}
	\setlength{\tabcolsep}{3pt}
	\begin{table}[tbp]
		\begin{center}
			\begin{tabular}{|c|c c c c c c c c c c c c c|}
				\hline
				&$L$ & $e_R$	&$\mu_R$&$\tau_R$	&$\nr$&$\edit\rho$&$\chi$&$\ssa$&$\ssb$&$S$&$\phia$&$\phib$&$\Delta$\\
				\hline
				$S_4$ &$\edit \rtp$ & $\rs$	&$\rs$&$\rs$&$\edit \rtp$&$\edit\rs$&$\edit \rtp$&$\rs$&$\rs$&$\rs$&$\rt$&$\rt$&$\rs$\\
				$\!\!C_4\!\!\times\!\!C_3\!\!$ &$i$ & $i$ &$i \om$&$i \ob$&$i$&$\edit 1$&$\om$&$1$&$1$&$-1$&$1$&$1$&$-1$\\
$\edit C_2$ &$\edit -1$ & $\edit 1$ &$\edit 1$&$\edit -1$&$\edit 1$&$\edit -1$&$\edit 1$&$\edit 1$&$\edit 1$&$\edit 1$&$\edit 1$&$\edit 1$&$\edit 1$\\
				$Y_{24}$ &$\rs$ & $\rs$	&$\rs$&$\rs$&$\rs$&$\edit \rs$&$\rs$&$\rs$&$\rs$&$\rs$&$\rx$&$\rx$&$\rxs_1$\\
				$C_2$ &$1$& $1$&$1$&$1$&$1$&$\edit 1$&$1$&$1$&$1$&$1$&$*$&$*$&$*$\\
				$C_2$ &$1$&$1$&$1$&$1$&$1$&$\edit1$&$1$&$-1$&$1$&$-1$&$-1$&$1$&$-1$\\
				$C_2$ &$1$& $1$&$1$&$1$&$1$&$\edit1$&$1$&$1$&$-1$&$-1$&$1$&$-1$&$-1$\\
				\hline
			\end{tabular}
		\end{center}
		\caption{The fields in the recast model as the multiplets under $(S_4\times C_4\times C_3 {\edit\, \times \, C_2})\times (Y_{24}\rtimes C_2 \times C_2 \times C_2)$.}
		\label{tab:flavourcontentnew}
\end{table}}

The field content of the model recast in the framework of the auxiliary group is given in Table~\ref{tab:flavourcontentnew}. The left-handed lepton field ($L$), the right-handed neutrino field ($N$) and the flavon field $\chi$ are triplets ($\rt$) under $S_4$ and invariants under $\bt$. The flavons $\phia$ and $\phib$ are not only triplets ($\rt$) under $S_4$, but also sextets ($\rx$) under $\bt$. $\Delta$ transforms as $\rxs_1$ under $\bt$. We impose the complex conjugation symmetry, $\phia\rightarrow\phia^*$, $\phib\rightarrow\phib^*$ and $\bind\rightarrow\bind^*$, represented as the $C_ 2$ in the 5\textsuperscript{th} row in Table~\ref{tab:flavourcontentnew}. Since $Y_{24}$ is a complex group, combining it with conjugation leads to the semidirect product, $Y_{24}\rtimes C_2$. {\edit Note that this conjugation symmetry is different from CP symmetry which involves the transformation of not only $\phia$, $\phib$ and $\bind$ but also all other complex scalars as well fermions in the model.} Two more $C_2$ groups are introduced (the last two rows in the table), so that $\ssa$, $\ssb$ and $S$ (similarly $\phia$, $\phib$ and $\Delta$) couple together in the Lagrangian.  The fermions, $L$, $e_R$, $\mu_R$, $\tau_R$ and $N$ are invariant under $Y_{24}\rtimes C_2 \times C_2 \times C_2$, hence it forms the auxiliary group, $G_x$. The rest of the flavour group ($S_4\times C_4\times C_3\,{\edit \times\,C_2 }$), which also appears in Table~\ref{tab:flavourcontent}, forms $G_r$.

Based on our field content and their symmetries, we construct the following Lagrangian:
\begin{align}\label{eq:lagrnew}
\begin{split}
{\mathcal L}=&y_\tau \bar{L} \frac{\chi}{\Lambda} \tau_R H+{\edit y_\mu \bar{L} \frac{\rho \chi^*}{\Lambda^2} \mu_R H + y_e \bar{L} \frac{\rho(\chi^*\chi)_{\rtp}}{\Lambda^3} e_R H} \\
& {+\edit y_\nu \bar{L} \nr \frac{\rho}{\Lambda}\widetilde{H} + y_s \left( \bar{\nr^c} \nr \right)_{\rs}  \frac{1}{\Lambda^2}\ssa S \ssb}\\
& {\edit + y_\phi \left( \bar{\nr^c} \nr \right)_{\rtp}    \frac{1}{\Lambda^2}\left[\left( \phia\bind\phib \right)_{\rtp}+\left( \phia^*\bind^*\phib^* \right)_{\rtp}\right].}
\end{split}
\end{align}
{\edit As was done in our original Lagrangian, Eq.~(\ref{eq:lagr}), we impose CP symmetry in the above Lagrangian, leading to real coupling constants, $y_x$.\edit} In Eq.~(\ref{eq:lagrnew}), the elementary multiplets $\phia$, $\phib$ and $\Delta$ couple together resulting in the effective triplet ($\rtp$) of $S_4$,
\be\label{eq:effective}
\edit \frac{1}{\Lambda^2}\left[\left( \phia\bind\phib \right)_{\rtp}+\left( \phia^*\bind^*\phib^* \right)_{\rtp}\right].
\ee
Eq.~(\ref{eq:lagrnew}) is the same as Eq.~(\ref{eq:lagr}), except for the fact that the triplet $\phi$ is replaced with the effective triplet and the singlet $s$ is replaced with the effective singlet $\edit \frac{1}{\Lambda^2}\ssa S \ssb$~\footnote{The only purpose of introducing the singlets $\ssa$ and $\ssb$ in the Lagrangian is to ensure that the effective singlet term has the same dimensionality as the effective triplet term.}.

The various components of the flavons $\Delta$, $\phia$ and $\phib$ can be expressed as $\bind_{\alpha}$, $\phia_{\beta m}$ and $\phib_{\gamma n}$ where the Greek and the Latin indices correspond to $\bt$ and $S_4$ respectively. To explicitly calculate the expression of the effective triplet, Eq.~(\ref{eq:effective}),  in terms of the components of the elementary flavons, we use the tensor product expansions given in Eqs.~(\ref{eq:tp3p}, \ref{eq:61}). Thus we obtain
\be\label{eq:cgcombined}
\left( \phia\bind\phib \right)_{\rtp l} =   \sum C'_{lmn} C_{\alpha \beta \gamma} \phia_{\beta m}\phib_{\gamma n} \bind_{\alpha},
\ee
where the summation is over all the repeated indices and 
\be\label{eq:cg1}
\begin{split}
	C'_{lmn}&= 1 \quad \forall \,\,\, l \neq m \neq n\\
	&=0 \quad \text{otherwise}
\end{split}
\ee
and 
\be\label{eq:cg2}
\begin{split}
	C_{lmn}&= 1 \,\,\, \forall \,\,\, (l~\text{mod}~3) \neq (m~\text{mod}~3) \neq (n~\text{mod}~3) \quad \& \\
	& \quad\quad\,\,\,\,\,\, (l-1)~\text{div}~3 = (m-1)~ \text{div}~3 = (n-1)~\text{div}~3 \\
	&=0 \,\,\, \text{otherwise}
\end{split}
\ee
are the C-G coefficients corresponding to Eq.~(\ref{eq:tp3p}) and Eq.~(\ref{eq:61}) respectively. In Eq.~(\ref{eq:cg2}), `mod' denotes the modulo operation and `div' denotes the integer division. $\left( \phia^*\bind^*\phib^* \right)_{\rtp}$ \!is simply the complex conjugate of $\left( \phia\bind\phib \right)_{\rtp} $.

The elementary flavons are assigned the following VEVs:
\begin{align}
\langle \phia \rangle &= v_\Phi
\left(\begin{matrix}
1 & 0 & 0\\
0 & \ta & 0\\
0 & 0 & 1\\
-i \om & 0 & 0\\
0 & \tb & 0\\
0 & 0 & i \ob
\end{matrix}\right), \quad \langle \phib \rangle =v_\Phi
\left(\begin{matrix}
i \om & 0 & 0\\
0 & \ta & 0\\
0 & 0 & -i \ob\\
-1 & 0 & 0\\
0 & \tb & 0\\
0 & 0 & -1
\end{matrix}\right),\label{eq:compflavvevs1}\\
\langle \bind \rangle & = v_\Delta
\left(\begin{matrix}
1 & 1& 1& 1& 1& 1
\end{matrix}\right),\label{eq:compflavvevs2}
\end{align}
where $v_\Phi$ and $v_\Delta$ are real. Here, $\langle \phia\rangle$ and $\langle \phib\rangle$ are given in matrix forms with the rows (columns) representing $\bt$ ($S_4$). Using Eqs.~(\ref{eq:cgcombined}, \ref{eq:compflavvevs1}, \ref{eq:compflavvevs2}), we obtain the vacuum alignment of the effective triplet, Eq.~(\ref{eq:effective}),
\be
\begin{split}
	&\edit \frac{1}{\Lambda^2}\left\langle\left( \phia\bind\phib \right)_{\rtp} + \left( \phia^*\bind^*\phib^* \right)_{\rtp}\right\rangle \\&\quad \quad\quad\quad\quad\quad{\edit =\frac{4v_\Phi^2v_\Delta}{\Lambda^2} \left(-\frac{1}{2\sqrt{2}}, -\sqrt{3},\frac{1}{2\sqrt{2}}\right).}
\end{split}
\ee
This alignment is proportional to the VEV of the triplet flavon that we originally proposed in Eq.~(\ref{eq:vevphi}). We also assign
\be
\langle \ssa \rangle = \langle \ssb \rangle = \langle S \rangle = v_S
\ee
so that 
\be
\edit \frac{1}{\Lambda^2}\left\langle \ssa S \ssb\right\rangle = \frac{v_S^3}{\Lambda^2}
\ee
which corresponds to the VEV of the singlet flavon, Eq.~(\ref{eq:vevs})\footnote{We did not include the $S_4$ effective singlet, $\frac{1}{\Lambda^2}\left[\left( \phia\bind\phib \right)_{\rs} + \left( \phia^*\bind^*\phib^* \right)_{\rs}\right]$, and the doublet, $\frac{1}{\Lambda^2}\left[\left( \phia\bind\phib \right)_{\rd} + \left( \phia^*\bind^*\phib^* \right)_{\rd}\right]$, in the Lagrangian, Eq.~(\ref{eq:lagrnew}), because they vanish for the given set of VEVs, Eqs.~(\ref{eq:compflavvevs1}, \ref{eq:compflavvevs2}).}. {\edit We noted in Section~4 that the VEVs of the flavons $\phi$ and $s$ are of the order of $10^{11}~\text{GeV}$. In the recast model, we have replaced these flavons with the corresponding effective multiplets. If we assume that the VEVs of all flavons, say $v_x$, in the recast model are at the same scale, we obtain $v_x\approx10^{15}~\text{GeV}$ and $\Lambda\approx10^{17}~\text{GeV}$.}

\subsection{The Residual Symmetries of $\langle\phia\rangle$, $\langle\phib\rangle$ and $\langle \Delta \rangle$}

In this section, we show that each of the VEVs, $\langle\phia\rangle$, $\langle\phib\rangle$, $\langle \Delta \rangle$, is an invariant under the action of certain elements of the flavour group. The VEV does not fully break the flavour group, rather it breaks the group into a specific subgroup that constitutes its residual symmetries. In this way, the alignment of each of these VEVs is uniquely defined. 

Let us first study the symmetries of the alignment 
\be\label{eq:vevmod}
\langle\Phi \rangle = |\langle\phia\rangle| = |\langle\phib\rangle| = v_\Phi \left(\begin{matrix}
	1 & 0 & 0\\
	0 & 1 & 0\\
	0 & 0 & 1\\
	1 & 0 & 0\\
	0 & 1 & 0\\
	0 & 0 & 1
\end{matrix}\right),
\ee
where $v_\Phi$ is real. This alignment remains invariant under the following group actions:
\begin{align}
\mathcal {O}_{1} \langle\Phi \rangle &= \Cma D \Cma D \,\, \langle\Phi \rangle \,\, QRQPQ \,\, = \langle\Phi\rangle,\label{eq:op1}\\
\mathcal {O}_{2} \langle\Phi \rangle &=\Cmb D \Cmb D \,\, \langle\Phi \rangle \,\, RQPQ^2 \,\, = \langle\Phi\rangle,\label{eq:op2}\\
\mathcal {O}_{3} \langle\Phi \rangle &=D \,\, \langle\Phi \rangle \,\, I= \langle\Phi\rangle,\label{eq:op3}\\
\mathcal {O}_{4} \langle\Phi \rangle &=E \,\, \langle\Phi \rangle \,\, Q \,\, = \langle\Phi\rangle,\label{eq:op4}\\
\mathcal {O}_{5} \langle\Phi \rangle &=\langle\Phi \rangle^* = \langle\Phi\rangle,\label{eq:op5}
\end{align}
where
\begin{align}
\Cma D \Cma D  &= \text{diag}(1,-1,-1,1,-1,-1),\\
QRQPQ &= \text{diag}(1, -1, -1),\\
\Cmb D \Cmb D  &= \text{diag}(-1,1,-1,-1,1,-1),\\
RQPQ^2 &= \text{diag}(-1, 1, -1).
\end{align}
The group actions $\mathcal {O}_{1}$ and $\mathcal {O}_{2}$, Eqs.~(\ref{eq:op1}, \ref{eq:op2}), multiply certain components of $\langle\Phi \rangle$ with $-1$. Invariance under this action ensures that those components vanish, i.e., Eqs.~(\ref{eq:op1}, \ref{eq:op2}) ensure that all the components marked with zeros in Eq.~(\ref{eq:vevmod}) vanish. $\mathcal {O}_{3}$ exchanges the first three rows of $\langle\Phi \rangle$ with the last three. Invariance under $\mathcal {O}_{3}$ ensures that the upper three rows are equal to the lower three. This condition is satisfied by $\langle\Phi \rangle$. The group action $\mathcal {O}_{4}$ cycles the following sets of components: \{(1,1), (2,2), (3,3)\}, \{(1,3), (2,1), (3,2)\}, \{(1,2), (2,3), (3,1)\}, \{(4,1), (5,2), (6,3)\}, \{(4,3), (5,1), (6,2)\}, \{(4,2), (5,3), (6,1)\}. Invariance under this action ensures that the components within each set are equal. This condition is also satisfied by the VEV. Finally, the invariance under the action of $\mathcal {O}_{5}$ ensures that the VEV is real. The conditions, Eqs.~(\ref{eq:op1}, \ref{eq:op2}, \ref{eq:op3}, \ref{eq:op4}, \ref{eq:op5}) uniquely lead to the required vacuum alignment, Eq.~(\ref{eq:vevmod}) up to a real proportionality constant. 

$\mathcal {O}_{1}$, $\mathcal {O}_{2}$, $\mathcal {O}_{3}$ and $\mathcal {O}_{5}$ generate various $C_2$ subgroups and $\mathcal {O}_{4}$ generates a $C_3$ subgroup of the flavour group. It can be shown that the group generated by $\mathcal {O}_{1}$, $\mathcal {O}_{2}$, $\mathcal {O}_{3}$, $\mathcal {O}_{4}$ and $\mathcal {O}_{5}$ has 192 elements and has the structure $(C_2 \times C_2 \times C_2 \times C_2)\rtimes C_3\times C_2 \times C_2$. This group represents the residual symmetries of $\langle\Phi \rangle$ and it uniquely defines $\langle\Phi \rangle$. 

$\langle\phia \rangle$ and $\langle\phib \rangle$, Eqs.~(\ref{eq:compflavvevs1}), can also be defined in terms of specific $(C_2 \times C_2 \times C_2 \times C_2)\rtimes C_3\times C_2 \times C_2$ subgroups of the flavour group. Consider the group elements
\begin{align}
\acute{g}&= \Cwb^2 \Ctb = \text{diag}(1, \ta, 1, -i \om, \tb, i \ob)\\
\grave{g}&= \Cmb D \Cwb^2 \Ctb^7 D= \text{diag}(i \om, \ta, -i \ob, -1, \tb, -1)
\end{align}
We may obtain $\langle\phia\rangle$ and $\langle\phib\rangle$ by the action of $\acute{g}$ and $\grave{g}$ respectively on $\langle\Phi \rangle$,
\be
\acute{g}\langle\Phi \rangle=\langle\phia\rangle,\quad \grave{g} \langle\Phi\rangle =\langle\phib\rangle.
\ee
We define a new set of group elements,
\be
\acute{\mathcal {O}}_{i} = \acute{g} \mathcal {O}_{i}  \acute{g}^{-1},\quad \grave{\mathcal {O}}_{i} = \grave{g} \mathcal {O}_{i}  \grave{g}^{-1},
\ee
where $i=1,..,5$. These elements uniquely define $\langle\phia\rangle$ and $\langle\phib\rangle$ in terms of their residual symmetries,
\be
\acute{\mathcal {O}}_{i} \langle\phia \rangle = \langle\phia \rangle, \quad \grave{\mathcal {O}}_{i} \langle\phib \rangle = \langle\phib \rangle.
\ee
$\langle\phia \rangle$ and $\langle\phib \rangle$, Eqs.~(\ref{eq:compflavvevs1}), uniquely break the flavour group into separate $(C_2 \times C_2 \times C_2 \times C_2)\rtimes C_3\times C_2 \times C_2$ groups generated by $\acute{\mathcal {O}}_{i}$ and $\grave{\mathcal {O}}_{i}$ respectively. $\langle\Phi\rangle$, $\langle\phia \rangle$ and $\langle\phib \rangle$ belong to the same orbit and the $(C_2 \times C_2 \times C_2 \times C_2)\rtimes C_3\times C_2 \times C_2$ groups generated by $\mathcal {O}_{i}$, $\acute{\mathcal {O}}_{i}$ and $\grave{\mathcal {O}}_{i}$ form their respective stabilisers.

The flavon $\Delta$ transforms as $\rxs_1$. The generators of $\rx_1$ are given in Eqs.~(\ref{eq:gensixone}). Using these generators we can show that $D(\rxs_1) = D(\rx) = D$ and $E(\rxs_1) = E(\rx) = E$. The vacuum alignment, $\langle\Delta \rangle = v_\Delta(1,1,1,1,1,1)$, Eq.~(\ref{eq:compflavvevs2}), remains invariant under the following group actions:
\be\label{eq:deltainvariance}
D \langle\Delta \rangle = \langle\Delta \rangle, \quad E \langle\Delta \rangle= \langle\Delta \rangle, \quad \langle\Delta \rangle^*= \langle\Delta \rangle.
\ee
$D$ interchanges the first three components of $\langle\Delta \rangle$ with the last three. Invariance under this action ensures that the $1^\text{st}$, the $2^\text{nd}$ and the $3^\text{rd}$ components are equal to the $4^\text{th}$, the $5^\text{th}$ and the $6^\text{th}$ components respectively. $E$ cycles the first three components as well as the last three. Invariance under $E$ ensures that the first three components are equal and also the last three components are equal. Invariance under both $D$ and $E$ implies that all the elements of the sextet are equal which is true for $\langle\Delta \rangle$. The third condition, $\langle\Delta \rangle^*= \langle\Delta \rangle$, implies that the VEV is real. Eqs.~(\ref{eq:deltainvariance}) uniquely results in the required VEV, Eq.~(\ref{eq:compflavvevs2}). $D$ and $E$ generate $C_2$ and $C_3$ respectively and together they generate $C_6$. We also have a $C_2$ symmetry representing the conjugation. Therefore, $\langle\Delta \rangle$ has the residual symmetry, $C_6\times C_2$. This residual symmetry uniquely defines $\langle\Delta \rangle$. 

{\edit Construction of the most general potential for the flavons presented in this paper would be quite tedious because of the very large number of degrees of freedom involved. We would also need driving fields to ensure that accidental Lie symmetries of the renormalizable potential are broken to obtain the required finite symmetries. Such details are beyond the scope of this paper. However, it is known~\cite{2011.11653,MICHEL200111,MICHEL1971} that for the potential constructed with an irreducible multiplet, every alignment that is fully determined by its residual symmetries is guaranteed to form a stationary point. In our model, VEVs of all irreducible multiplets are fully determined by their respective residual symmetries. Therefore, the generation of these VEVs is automatically ensured.} Undesirable cross terms among the various multiplets can be avoided by invoking mechanisms that are widely assumed in flavour models such as localisation in branes in an extradimensional framework. 

The residual symmetries that define the VEVs of the individual elementary multiplets $\phia$, $\phib$ and $\Delta$ may not survive when the flavons are taken together to form the effective triplet. On the other hand, we utilise the fact that the flavon VEVs are oriented along different directions (with different residual symmetries) to obtain the required VEV of the effective triplet.  Even though the effective triplet is an invariant in the space of the auxiliary group, it is constructed from the multiplets ($\phia$, $\phib$ and $\Delta$) transforming non-trivially in that space. This implies that the components of the effective triplet hold information about how these multiplets are oriented relative to each other\footnote{As an analogy, consider the scalar product of two vectors. It is nothing but an invariant constructed from the vectors and it holds information about their relative orientation.}. We may recall that the second half of this paper is simply the pursuit of finding an explanation for the angle $\alpha=-\frac{\pi}{6}$ proposed for the triplet VEV, Eq.~(\ref{eq:vevalpha}). The relative orientations of the VEVs of $\phia$, $\phib$ and $\Delta$ in the higher dimensional flavour space (uniquely defined by their respective residual symmetries) contains precisely the information about this angle, which in turn gets transferred to the VEV of the effective triplet living in the lower dimensional space of the $S_4$ triplet representation. 

{\edit The order of the symmetry group in the model is extremely large ($>10^6$). This can be viewed upon as an unpleasant feature of the framework of the auxiliary group. In this context, we may note that Ref.~\cite{Babu:2010bx} used a group constructed as a wreath product (a special case of the semidirect product) with order $>10^4$ to decouple the flavon VEVs in the charged-lepton and neutrino sectors so that tribimaximal mixing (TBM) can be obtained. We may contrast this with the size of $A_4$ group (order $12$), which is widely used for constructing TBM even though it results in dangerous cross terms that spoil the required VEVs. As stated earlier, model builders forbid these terms with the help of assumptions like extra dimensions. Having a larger flavour group, as in Ref.~\cite{Babu:2010bx}, can be considered as a necessary trade-off to avoid such assumptions. A simpler semidirect product, $Q_8\rtimes A_4$, which is smaller than the wreath product, was later proposed~\cite{Holthausen:2011vd,Holthausen:2012wz} to achieve decoupling and obtain TBM. In the current paper, the flavour group is given as the direct product, $G_f=G_r\times G_x$. We may be able to reduce the size of $G_f$ by a couple of orders of magnitude if we replace the direct product with a suitable semidirect product. The presence of a large number of zeros in the VEVs of the elementary multiplets, Eqs.~(\ref{eq:compflavvevs1}), is a hint towards this. 

Unlike Refs.\cite{Babu:2010bx,Holthausen:2011vd,Holthausen:2012wz} where the purpose of using an enlarged group was to decouple VEVs, the current paper uses it to construct a novel VEV for the effective multiplet, which in turn generates a small reactor angle $\sin^2\theta_{13}=\frac{1}{3}\sin^2 \frac{\pi}{12}$. We constructed the group $Y_{24}$ (the main constituent of our auxiliary group) as a six-dimensional representation containing phases $e^{i\frac{\pi}{12}}$. Such constructions are rather unavoidable given our philosophy of modelling small mixing angles using finite groups. Consequently, we obtain very large groups. Nevertheless, we argue that rather than looking at flavor groups from the perspective of their size, it would be instructive to study the nature of their generators. Even though the group $Y_{24}$ is very large, its generators are fairly simple, Eqs.~(\ref{eq:ygens},\ref{eq:ygens1}). They correspond to certain permutations and diagonal multiplications with complex phases. If there is an underlying mechanism behind the origin of these generators, we may expect it to be simple.}

\section{Conclusion}
\label{sec:conclusion}

$\text{TM}_1$ mixing preserves the first column of the tribimaximal mixing. $\text{TM}_1$ can be parametrised using an angle ($\theta$) and a phase ($\zeta$). In this parametrisation, the reactor mixing angle is given by $\sin^2\theta_{13}=\frac{1}{3}\sin^2 \theta$. In this paper, we construct a type-1 seesaw model based on the flavour group, $S_4\times C_4\times C_3$,  and obtain $\text{TM}_1$ mixing with $\theta = -\frac{\pi}{12}$ and $\zeta = \frac{\pi}{2}$ consistent with the current experimental data. The mixing obtained exhibits $\mu$-$\tau$ reflection symmetry so that we have $\theta_{23}=\frac{\pi}{4}$. We also obtain maximal CP violation with $\delta = -\frac{\pi}{2}$. The light neutrino masses are found to be functions of two model parameters and by fitting them with the experimental values of the mass-squared-differences we predict the individual masses. They follow the normal ordering. We also predict the effective neutrino mass applicable to the neutrinoless double-beta decay. 

To construct the fermion mass terms, we introduce the flavons $\chi$, $s$, $\phi$ which transform as $\rt$, $\rs$ and $\rtp$ respectively under $S_4$. We construct the flavon potentials which when minimised lead to their vacuum expectation values. The flavon $\chi$ couples in the charged-lepton sector and its vacuum alignment leads to the $3\times 3$-trimaximal contribution to the neutrino mixing matrix. The flavons $s$ and $\phi$ couple in the Majorana sector. Their vacuum alignments constitute the Majorana mass matrix. The product of the $3\times 3$-trimaximal matrix and the diagonalising matrix from the Majorana sector generates the $\text{TM}_1$ mixing with $\theta = -\frac{\pi}{12}$ and $\zeta = \frac{\pi}{2}$. Obtaining the required vacuum alignment of $\phi$ entails tuning of the parameters in the flavon potential. This tuning implies that the symmetries of $S_4$ are not sufficient to uniquely determine this vacuum alignment. Therefore, we recast the model by introducing additional symmetries using the framework of the auxiliary group.

In this framework, the flavour group is given as the direct product of a `conventional' flavour group and an auxiliary group. Our original flavour group ($S_4\times C_4\times C_3{\edit \times C_2}$) takes the place of the conventional flavour group in the recast model. The auxiliary group is given by $Y_{24}\rtimes C_2 \times C_2 \times C_2$ where $Y_{24}$ is a discrete group that we constructed with the required symmetries. We introduce the flavons $\phia$, $\phib$ and $\Delta$ which transform under both the conventional flavour group and the auxiliary group. They are coupled together to obtain $\left( \phia\bind\phib \right)_{\rtp} +\left( \phia^*\bind^*\phib^* \right)_{\rtp}$ which is a triplet under $S_4$ and an invariant under the auxiliary group. This effective triplet transforms the same way as the triplet $\phi$ present in the original model. We assign specific VEVs to $\phia$, $\phib$ and $\Delta$ and identify the residual symmetries corresponding to each of them as a subgroup of the flavour group, i.e. ~a subgroup of $(S_4\times C_4\times C_3{\edit \times C_2})\times (Y_{24}\rtimes (C_2 \times C_2 \times C_2))$. We find that these subgroups uniquely define the VEVs. Using these VEVs, we uniquely obtain the VEV of the effective triplet also, which is in agreement with the VEV of $\phi$ proposed in the original model. 

\begin{acknowledgements}
I would like to thank Paul Harrison and Bill Scott for the stimulating discussions. I acknowledge the help and support from Ambar Ghosal and Debasish Majumdar. I am grateful to Sujatha Ramakrishnan for helping me with constructing the figures.
\end{acknowledgements}

\bibliographystyle{JHEP}
\bibliography{tm1piby12.bib, noninspire.bib}
\end{document}